\documentclass[aps,prb,reprint,amsmath,amssymb,groupedaddress]{revtex4-1}

\usepackage{graphicx}
\usepackage{dcolumn}
\usepackage{bm}
\usepackage{color}
\usepackage{hyperref}

\newcommand{\ve}[1]{\mathbf{#1}}

\newcommand{\exciting}{{\usefont{T1}{lmtt}{b}{n}exciting}}

\begin{document}

\title{Electronic and optical excitations at the pyridine/ZnO(10$\overline{1}$0) hybrid interface}

\author{Olga Turkina}
\affiliation{Institut f\"ur Physik and IRIS Adlershof, Humboldt-Universit\"at zu Berlin, Berlin, Germany}
\email{olga.turkina@physik.hu-berlin.de}
\author{Dmitrii Nabok}
\affiliation{Institut f\"ur Physik and IRIS Adlershof, Humboldt-Universit\"at zu Berlin, Berlin, Germany}
\author{Andris Gulans}
\affiliation{Institut f\"ur Physik and IRIS Adlershof, Humboldt-Universit\"at zu Berlin, Berlin, Germany}
\author{Caterina Cocchi}
\affiliation{Institut f\"ur Physik and IRIS Adlershof, Humboldt-Universit\"at zu Berlin, Berlin, Germany}
\author{Claudia Draxl}
\affiliation{Institut f\"ur Physik and IRIS Adlershof, Humboldt-Universit\"at zu Berlin, Berlin, Germany}

\date{\today}

\begin{abstract}
By combining all-electron density-functional theory with many-body perturbation theory, we investigate a prototypical inorganic/organic hybrid system, composed of pyridine molecules that are chemisorbed on the non-polar ZnO($10\overline{1}0$) surface. We employ the $G_0W_0$ approximation to describe its one-particle excitations in terms of the quasi-particle band structure, and solve the Bethe-Salpeter equation for obtaining the absorption spectrum. The different character of the constituents leads to very diverse self-energy corrections of individual Kohn-Sham states, and thus the $G_0W_0$ band structure is distinctively different from its DFT counterpart, i.e., many-body effects cannot be regarded as a rigid shift of the conduction bands. We explore the nature of the optical excitations at the interface over a wide energy range and show that various kinds of electron-hole pairs are formed, comprising hybrid excitons and (hybrid) charge-transfer excitations. The absorption onset is characterized by a strongly bound bright ZnO-dominated hybrid exciton. For selected examples of either exciton type, we analyze the individual contributions from the valence and conduction bands and discuss the binding strength and extension of the electron-hole wavefunctions. 
\end{abstract}


\maketitle

\section{\label{Introduction}Introduction}

Combining the best of two worlds is the main aim behind intensive investigations of hybrid materials. In organic/inorganic hybrid systems, for instance, their major advantages for opto-electronic applications are seen in the strong light-matter coupling of the organic components, while the inorganic counterparts excel in large carrier mobilities and thus efficient charge-carrier transport as well as small exciton binding energies.\cite{ishii1999energy,Hybrid_review,wright2012organic} These properties together with a type-I level alignment at the hybrid interface would be ideally suited for light-emitting applications.\cite{Coe,xu2011conjugated} Type-II alignment, in turn, is favored for a hybrid solar cell where optical absorption of sunlight is expected to directly lead to electron-hole separation.\cite{chan2012direct,nagata2013photoelectron,gao2013renaissance,bianchi2014cascade,piersimoni2015charge} Prerequisite for that is the light-induced creation of hybrid or charge-transfer excitons, which exhibit the hole to a large extent on one side of the interface while the excited electron would reside on the other side. Investigations along these lines on hybrid inorganic/organic materials composed of wide-gap semiconductors like ZnO and organic chromophores have created an active field of research.\cite{blumstengel2010band,wood2012binding,xu2013,Schlesinger2013,musselman2014improved,lange2014tuning,timpel2014surface,lange2015zinc,friede2015nanoscale,schlesinger2015efficient,timpel2015energy,kirmse2016structure,hofmann2017band,stahler2017global,wang2018dynamic} 

The adsorption of a molecular layer on inorganic surfaces has also been exploited to tune the work function of the inorganic component.\cite{hill1998molecular,Schlesinger2013,lange2014tuning,timpel2014surface} A special prototypical hybrid system, consisting of a monolayer of pyridine molecules chemisorbed on the non-polar ZnO($10\overline{1}0$) surface (labelled Py/ZnO in the following) has been studied experimentally by photoelectron spectroscopy to determine the reduction of the electron injection barrier at the interface.\cite{Hofmann}
The binding mechanism and the interface morphology of this system has been studied theoretically by density functional theory (DFT), including van der Waals interactions.\cite{Hofmann}

Insight from {\it ab initio} theory is essential in order to gain understanding of the opto-electronic properties of such complex materials. However, even when applying state-of-the-art methods, a fully quantatitive description of hybrid materials remains extremely challenging.\cite{DNH} Approximations most suitable for the organic part may not be applicable to the inorganic one, and vice versa.\cite{migani2014quasiparticle,liu2017energy} This situation may hamper obtaining the correct level alignment at the interface which, in turn, is the base for optical excitations. Further challenges are presented by the sensible interplay of geometry, bonding, hybridization, and electron-transfer processes. Ultimately, calculations of hybrid materials are computationally very demanding due to the large number of atoms per unit cell. Indeed, so far the electronic and optical properties of only a few prototypical hybrid systems have been successfully investigated by means of first-principles many-body approaches.\cite{milko2013evidence,mowbray2016optical,fu2017graphene,ljungberg2017charge}

In this work, we study Py/ZnO making use of the structure determined in Ref.~\onlinecite{Hofmann}. The $G_0W_0$ approximation of many-body perturbation theory (MBPT) is employed on top of DFT to obtain the level alignment at the interface, and the Bethe-Salpeter equation (BSE) is solved on the search for hybrid and charge transfer excitations. We demonstrate the role of many-body effects in both approaches, discussing the impact of the electron self-energy on the quasi-particle band structure and the existence of pronounced excitonic effects in the optical absorption spectra. We provide a detailed analysis of the absorption features, determining their character in terms of spatial extension and electron-hole binding strength. Finally, we critically evaluate the used methodology.

\begin{figure}[htb]
\includegraphics[width=0.50\columnwidth]{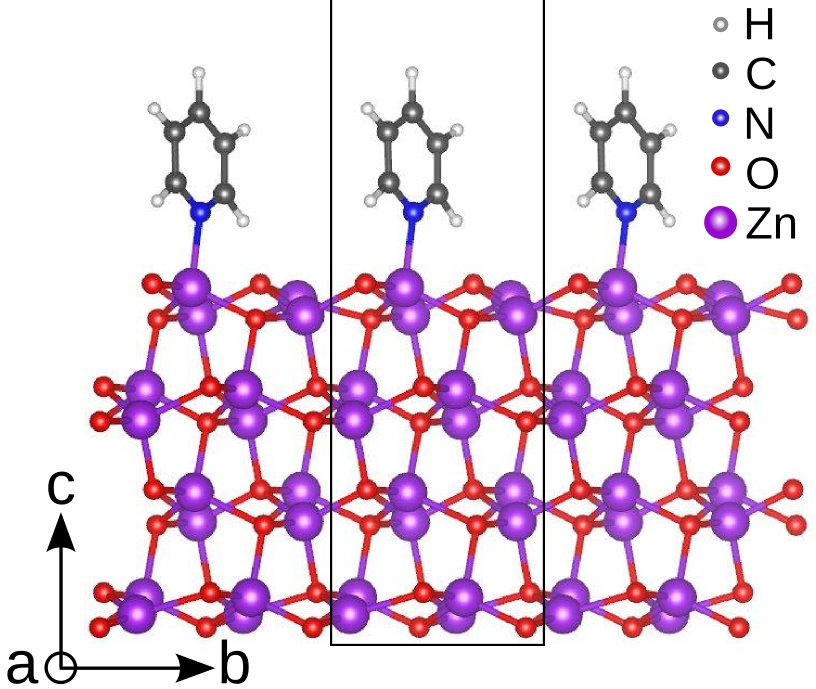}\hspace{0.04\columnwidth}
\includegraphics[width=0.44\columnwidth]{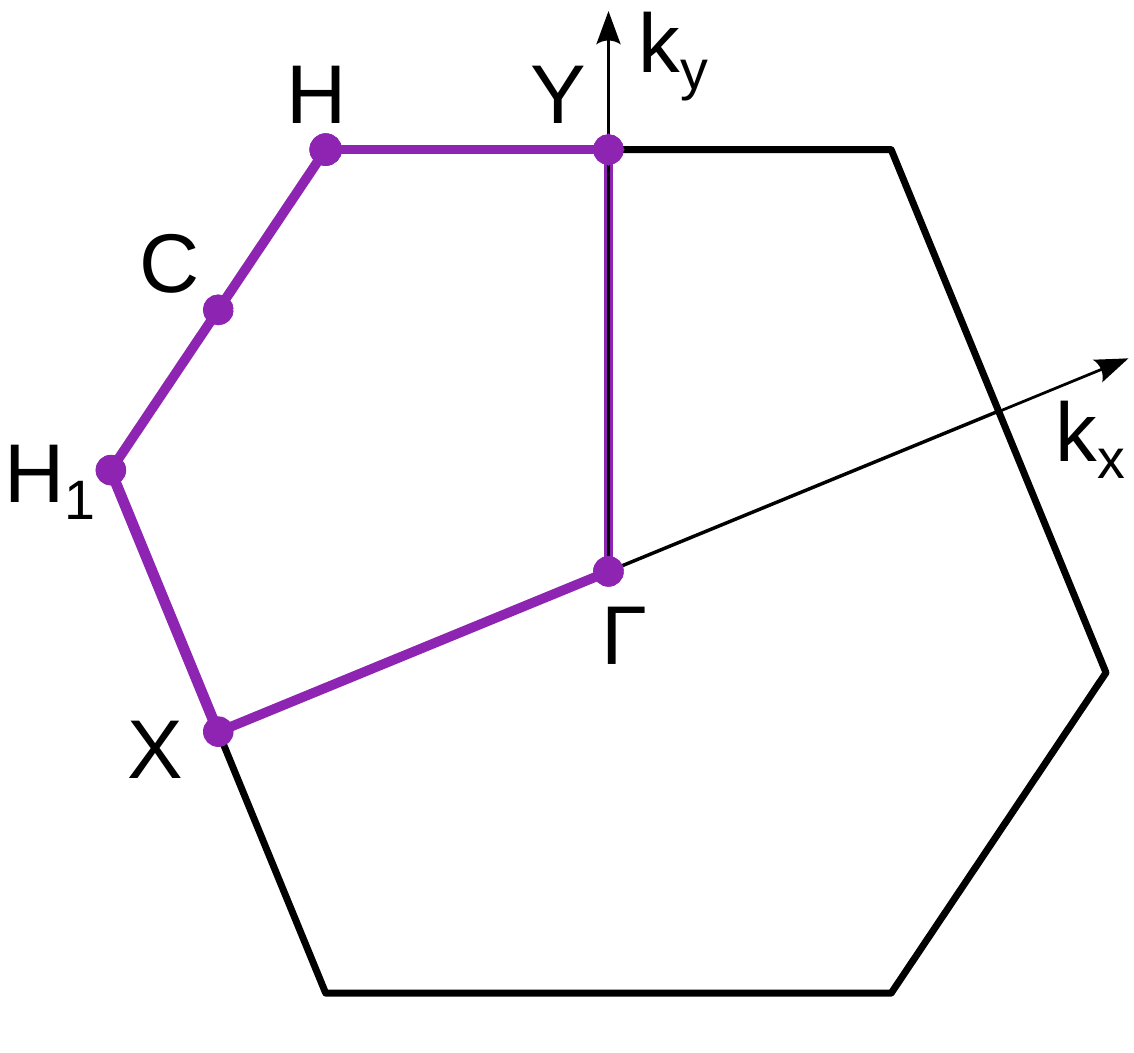}
\caption{\label{structure}Left: Side view of a pyridine monolayer on the non-polar ZnO(10$\overline{1}$0) surface. The unit cell is indicated by the black lines, with vacuum extending in c direction. Right: Surface Brillouin zone of the Py/ZnO(10$\overline{1}$0), with the band structure path marked in purple.}
\end{figure}

\section{\label{Methodology}Methodology}

\subsection{Theoretical background}

DFT is used to compute the ground-state properties, with the local-density approximation (Perdew-Wang parametrization\cite{LDA_PW,LDA_CA}) as the exchange-correlation functional (xc). Using the resulting Kohn-Sham (KS) eigenvalues and orbitals as a starting point, many-body perturbation theory is employed to determine the excited-state properties. The quasi-particle (QP) band structure is hereby calculated within the $G_0W_0$ approximation, providing the first order perturbative correction to the KS energies. The optical spectrum is obtained from the solution of the Bethe-Salpeter equation, an effective two-particle equation of motion for the electron-hole Green function. It is formulated as an eigenvalue problem as
\begin{equation}
 \sum_{v'c'\ve{k'}} H_{vc\ve{k},v'c'\ve{k'}} \: A_{v'c'\ve{k'}}^\lambda = E^\lambda A_{vc\ve{k}}^\lambda, 
\label{H_BSE}
\end{equation}
describing transitions from the valence ($v$) to the conduction ($c$) band region with the eigenvalues $E^\lambda$ representing the excitation energies. In the effective many-body Hamiltonian for singlet states, $H_{\rm BSE} = H_{\rm diag}+2H_{\rm x}+H_{\rm dir}$, the first term $H_{\rm diag}$ accounts for vertical transitions, while the other two terms consist of the electron-hole exchange $H_{\rm x}$ (not present for spin triplets) and the screened Coulomb interaction $H_{\rm dir}$. The attractive nature of the last term is responsible for the formation of bound excitons. Considering the first term only corresponds to the independent particle picture where the electron and hole do not interact. We will use this approximation for comparison in order to demonstrate the impact of the electron-hole interaction on the optical absorption.

The excitonic wavefunction is given by the linear combination of the KS wavefunctions, weighted by the coupling coefficients $A_{vc\ve{k}}^\lambda$, which are the eigenvectors in Eq. \ref{H_BSE}:
\begin{equation}\label{eh_wavefunction}
 \Phi^\lambda(\ve{r}_e,\ve{r}_h) = \sum_{vc\ve{k}} A_{vc\ve{k}}^\lambda \psi^*_{v\ve{k}}(\ve{r}_h) \psi_{c\ve{k}}(\ve{r}_e). 
\end{equation}
These eigenvectors enter the expression of the imaginary part of the macroscopic dielectric function which is used to represent the optical absorption spectrum
\begin{equation}
 \text{Im} \: \epsilon_M (\omega) = \frac{8 \pi^2}{\Omega} \sum_\lambda \left| \sum_{vc\ve{k}} A_{vc\ve{k}}^\lambda \frac{ \left\langle v\ve{k}| p_i|c\ve{k} \right\rangle }{ \epsilon_{c\ve{k}} - \epsilon_{v\ve{k}}} \right|^2 \delta(E^\lambda - \omega),
\end{equation}
and also yield information about the individual excitations in terms of the transition weights
\begin{equation}
 w_{ck}^\lambda = \sum_{v} |A_{vc\ve{k}}^\lambda|^2
 \label{eq:wc}
\end{equation}
and 
\begin{equation}
 w_{vk}^\lambda = \sum_{c} |A_{vc\ve{k}}^\lambda|^2,
 \label{eq:wv}
\end{equation}
which describe the composition of an excitation in terms of contributing bands. 

\subsection{Computational details}

The system considered in this work consists of four layers of ZnO covered by an upright-standing monolayer of pyridine molecules (Fig.~\ref{structure}). The geometry taken from Ref.~\onlinecite{Hofmann} has been slightly modified by including only one pyridine per unit cell, and thus including overall 43 atoms.
The original structure from Ref.~\onlinecite{Hofmann} contains two inequivalent molecules exhibiting a slightly different tilt angle with respect to the surface. 

All calculations are performed using the all-electron full-potential package \exciting,\ \cite{exciting} which is based on the linearized augmented planewave plus local-orbitals method. The adopted muffin-tin radii for the involved species are $R^{\rm H}_{\rm MT}= 0.8\, \text{bohr}$, $R^{\rm N}_{\rm MT}=R^{\rm C}_{\rm MT}= 1.2\, \text{bohr}$, $R^{\rm Zn}_{\rm MT}= R^{\rm O}_{\rm MT}= 1.6\, \text{bohr}$. A basis-set cutoff of $|\ve{G}+\ve{k}|_{max}=5$ for the smallest muffin-tin sphere is used. 8 $\text{\AA}$ of vacuum are included perpendicular to the surface to minimize the interactions between periodic slabs. The calculations of QP corrections to the KS eigenvalues within the $G_0W_0$ approximation include 1000 empty bands. The BZ sampling is performed with a $4 \times 4 \times 1$ $\ve{k}$ grid. Wannier interpolation is used to visualize the band structure and the density of states. The BSE is solved within the Tamm-Dancoff approximation. $150$ conduction bands are included in the calculation of the response function and the screened Coulomb potential. $41$ occupied and $25$ unoccupied bands on a $16 \times 16 \times 1$ $\ve{k}$ grid are taken into account in the construction of the BSE Hamiltonian. Local-field effects are considered by including 41 $\ve{G}+\ve{q}$ vectors. A Lorentzian broadening of 0.1 eV is included in the resulting optical spectra. Atomic structures and isosurfaces are obtained with the VESTA software.\cite{Vesta}

\section{Results}

\subsection{Electronic structure}

\begin{figure*}
\centering
\includegraphics[width=0.95\textwidth]{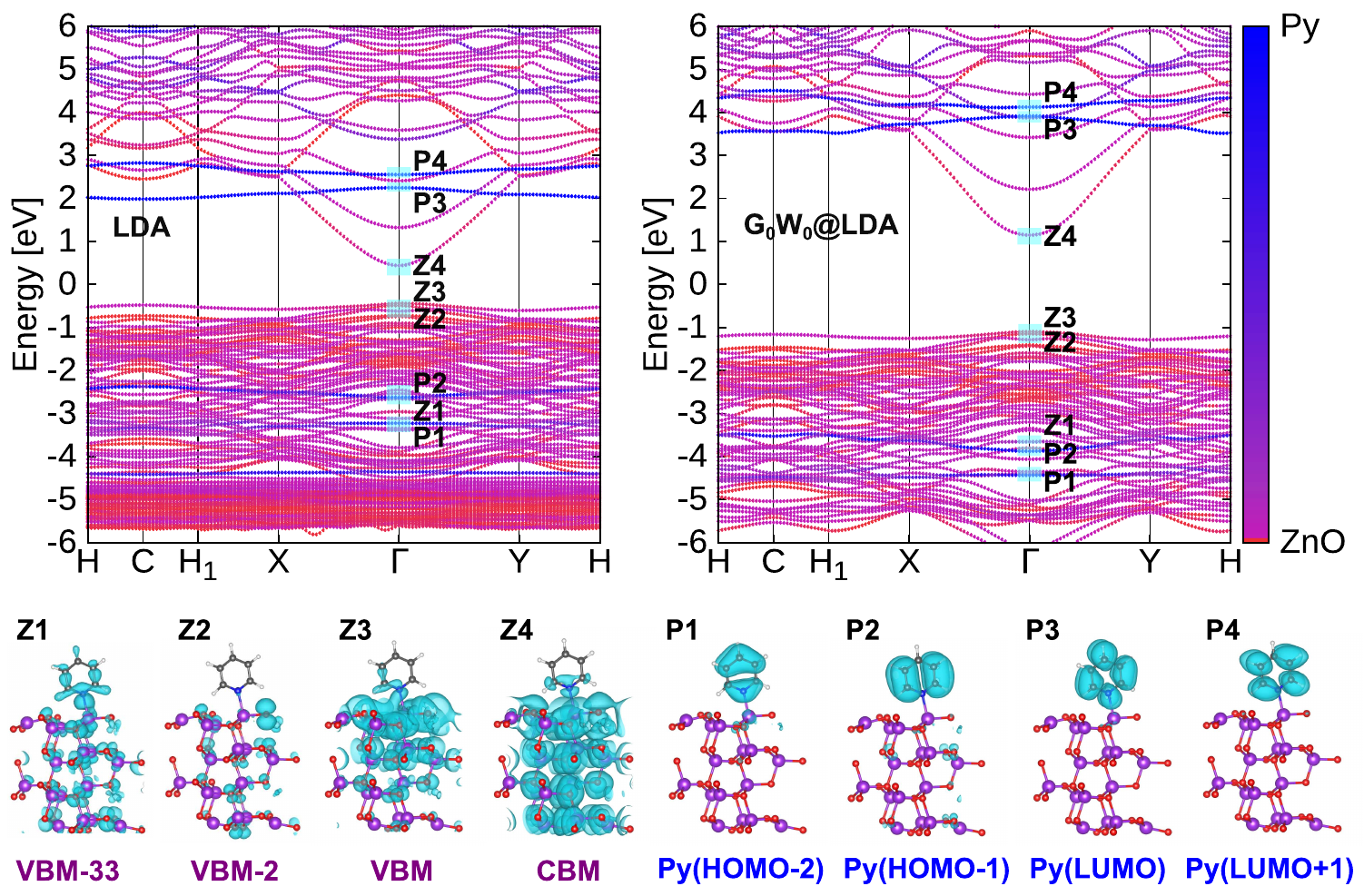}
\caption{\label{bandstructure}LDA (left) and QP band structure (right) of Py/ZnO. The color code indicates the band character, going from red (ZnO) over shades of purple (hybridized states) to blue (Py). The Fermi level is located at 0~eV. Bottom: Kohn-Sham wavefunctions of selected bands at the $\ve{\Gamma}$ point indicated in the panels above.}
\end{figure*}

The $G_0W_0$ band structure of the Py/ZnO interface is shown in Fig.~\ref{bandstructure} (top panel, right-hand side) together with the LDA results (left side) for comparison. The material exhibits a direct QP band gap of 2.25~eV at the $\Gamma$ point where the valence-band maximum (VBM) and the conduction-band minimum (CBm) have a predominant ZnO character, as highlighted by the color code of the bands (Fig.~\ref{bandstructure}). As such, the system can be regarded as a type-I heterostructure. The occupied (unoccupied) molecular orbitals of pyridine can be identified at lower (higher) energies in the valence (conduction) region. Supported by inspection of the wavefunctions, reported in the bottom part of Fig.~\ref{bandstructure}, we recognize the LUMO and LUMO+1 of the molecule, referred to in the following as Py(LUMO) and Py(LUMO+1), at about 2.5 eV and 3 eV with respect to the CBm, respectively. Likewise, in the valence band, Py(HOMO-1) and Py(HOMO-2) appear 2.5 eV and 3 eV below the VBM. Remarkably, the Py(HOMO) is no longer present as pure molecular state in the electronic structure of the hybrid interface.

A closer inspection of the wavefunctions of the frontier states, as reported in Fig.~\ref{bandstructure}, 
shows that the electronic structure of the Py/ZnO interface is more complex than what a simple model can suggest. The wavefunctions corresponding to the VBM and CBm at $\Gamma$ are, in fact, mostly delocalized over the ZnO surface. The CBm is a bulk-like state formed mainly by oxygen states whereas the VBM is an oxygen-dominated surface state. However, in both cases there is significant hybridization between the uppermost ZnO layer and the molecule along the Zn-N bond and towards the carbon ring, as discussed in detail in Ref.~\onlinecite{Hofmann}. {The HOMO of pyridine is now completely hybridized with ZnO bands and located in the upper part of the valence region. Deeper down in the valence band, we find more hybridized states where the electron distribution is equally spread in the inorganic and organic component. As an example of this type of states we show VBM-33 in Fig.~\ref{bandstructure}. All these features have relevant implications for the optical excitations of this system, which will be discussed in the next section.

The hybridization that occurs in the electronic structure is better highlighted in the density of states (DOS), as displayed in Fig.~\ref{dos} projected onto individual atoms (top panel) and the hybrid's constitutents (middle panel). For comparison also the DOS of the isolated sub-systems is shown (bottom panel). Although only minor contributions from pyridine appear in the uppermost part of the valence band,~\cite{Hofmann} there is clear indication of strong hybridization of the molecular states with the underlying ZnO surface at -3 eV and below. Here, the peaks associated with pyridine contributions are significantly broadened compared to the case of the isolated molecular monolayer. The positions of the (occupied and unoccupied) pyridine-derived bands in the hybrid system (middle panel) are shifted by about 1~eV towards each other, compared to the states of the monolayer (bottom panel). This effect, leading to a reduction of the molecular band gap, is known as the polarization-induced renormalization of molecular levels occuring in the presence of a substrate, and has been investigated at various metallic and semiconductor surfaces.\cite{Polarization1,Polarization2,Polarization3,Polarization4} 

We conclude this section by commenting about the adopted methodology to describe the electronic structure of this hybrid interface. While it is evident that the LDA functional underestimates the size of the band gap, yielding a value of 0.88~eV, the $G_0W_0$ approximation provides a more realistic result of 2.25~eV. Although this value, based on a semi-local xc functional as a starting point, is still expected to underestimate the true QP gap, the conclusions drawn above are unaffected. We emphasize that LDA was not only used to keep the calculations of such complex system feasible, but also for the fact that we do not have any xc functional in hand that would be particularly suited for hybrid materials. Thinking, \textit{e.g.}, on hybrid functionals, the organic side would require a much larger amount of Fock exchange than the inorganic counterpart. Recently, optimally tuned range-separated functionals have been proposed to improve the description of the electronic structure of hybrid interfaces with metallic substrates.~\cite{egger2015reliable,liu2017energy}
Hence, in any case, a fully quantitative discussion of such materials is still not possible.~\cite{DNH}

\begin{figure}
\centering
\includegraphics[width=0.90\columnwidth]{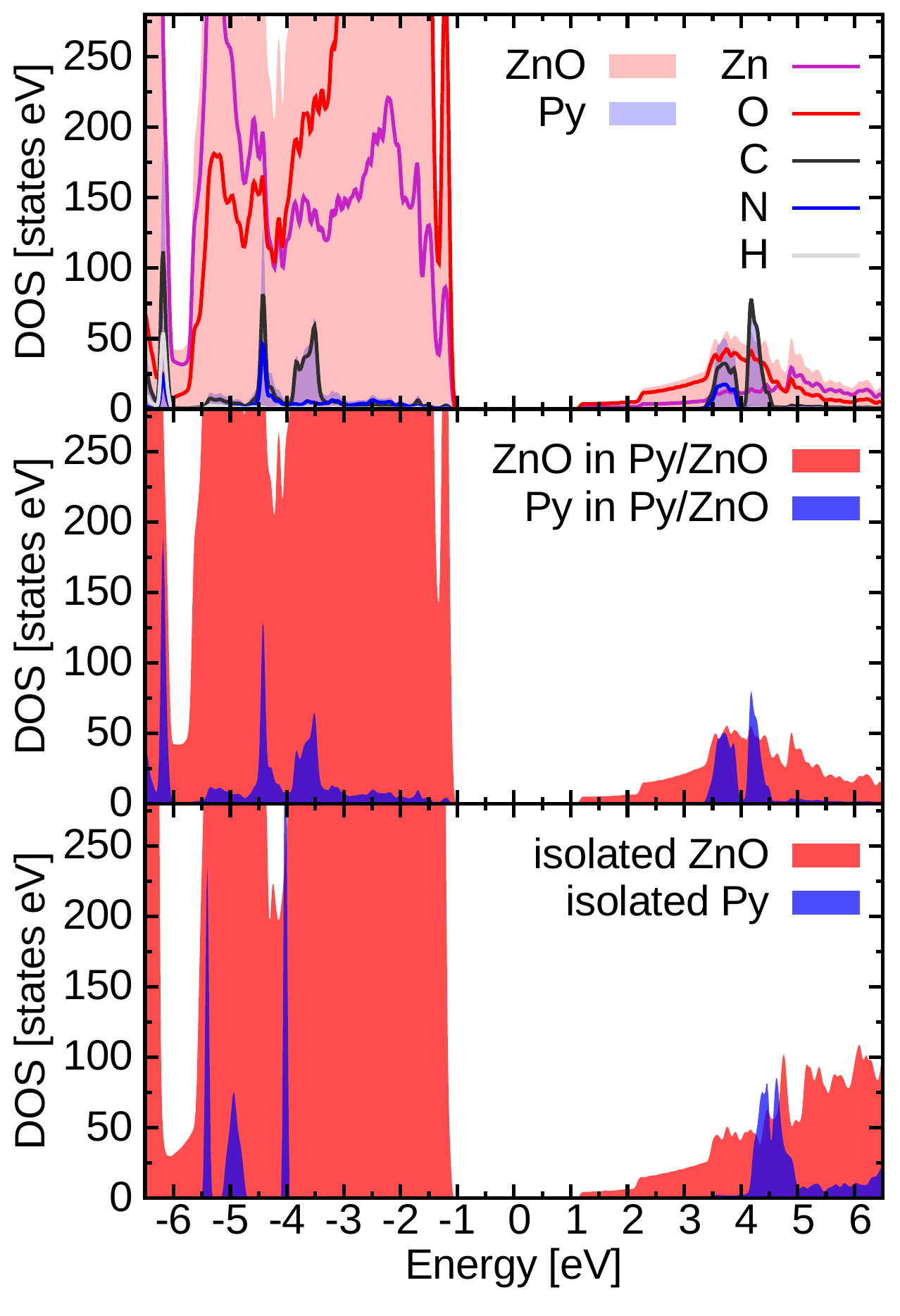}
\caption{\label{dos} Top: Atom-resolved QP density of states of Py/ZnO (lines) with the total contributions from pyridine and ZnO highlighted by the shaded area. Middle panel: Pyridine and ZnO contributions to the QP DOS. Bottom: QP DOS of isolated pyridine monolayer and ZnO slab as calculated in the same geometry as the hybrid system.}
\end{figure}

\subsection{Optical excitations}

\begin{figure}
\includegraphics[width=0.49\textwidth]{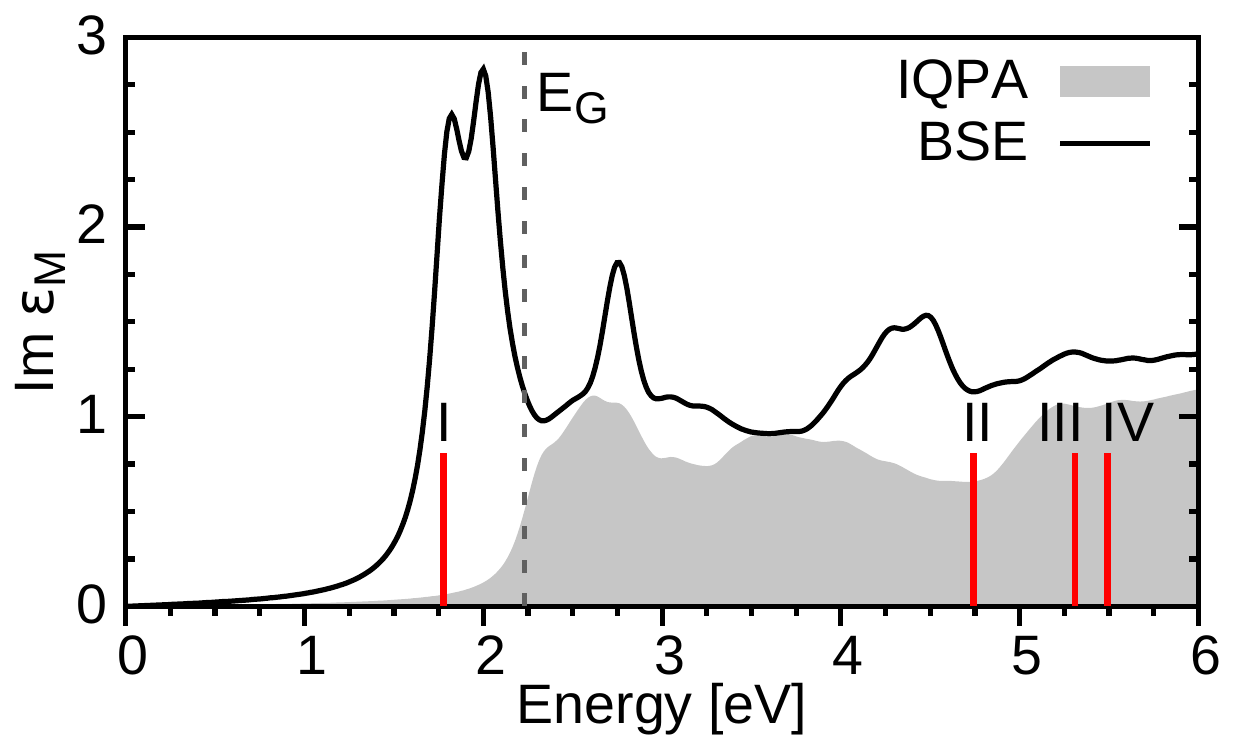}
\caption{Optical absorption spectrum of Py/ZnO averaged over the in-plane components of $\text{Im}\,\epsilon_M$ including (BSE, solid line) and neglecting (IQPA, grey area) excitonic effects. Red bars indicate the energetic position of selected excitons while the dashed line marks the position of the QP band gap.}
\label{spectrum}
\end{figure}

The complex electronic structure of the Py/ZnO interface, as discussed above, gives rise to a diversified spectrum of optical excitations. In Fig.~\ref{spectrum} we plot the absorption spectrum of this system as the imaginary part of the frequency-dependent macroscopic dielectric tensor, averaged over its in-plane components obtained from the solution of the BSE. The counterpart obtained in the independent QP approximation (IQPA), where the \textit{e-h} interaction is neglected, is shown for comparison. The absorption onset falls in the visible region and is dominated by two intense peaks below the band gap. These features, which are absent in the IQPA spectrum, are interpreted as bound excitons. The overall shape of the near-edge absorption resembles very much the profile of the dielectric function of bulk ZnO,~\cite{gori2010optical} where, however excitonic effects are much less pronounced. In fact, in the periodic crystal binding energies are of the order of 60 meV~\cite{gori2010optical} while here the first excitation (marked as I in Fig.~\ref{spectrum}) has a binding energy of about 0.4 eV.
Such enhancement of one order of magnitude is not surprising since the coupling strength depends on dielectric screening and the dimensionality of the system,\cite{Puschnig2002} and exciton formation in a ZnO surface has indeed been detected experimentally.~\cite{deinert2014ultrafast} The presence of only four layers further enhances quantum confinement effects that, in turn, lead to an increase of the binding strength.
In addition, the \textit{e-h} interaction tends to redistribute the spectral weight towards lower energies, such that the lowest-energy peaks are more intense compared to the above-edge region. We note in passing that the out of-plane component of the dielectric tensor is much smaller in magnitude and misses the first excitonic feature.

\begin{figure*}
	\includegraphics[width=0.47\textwidth]{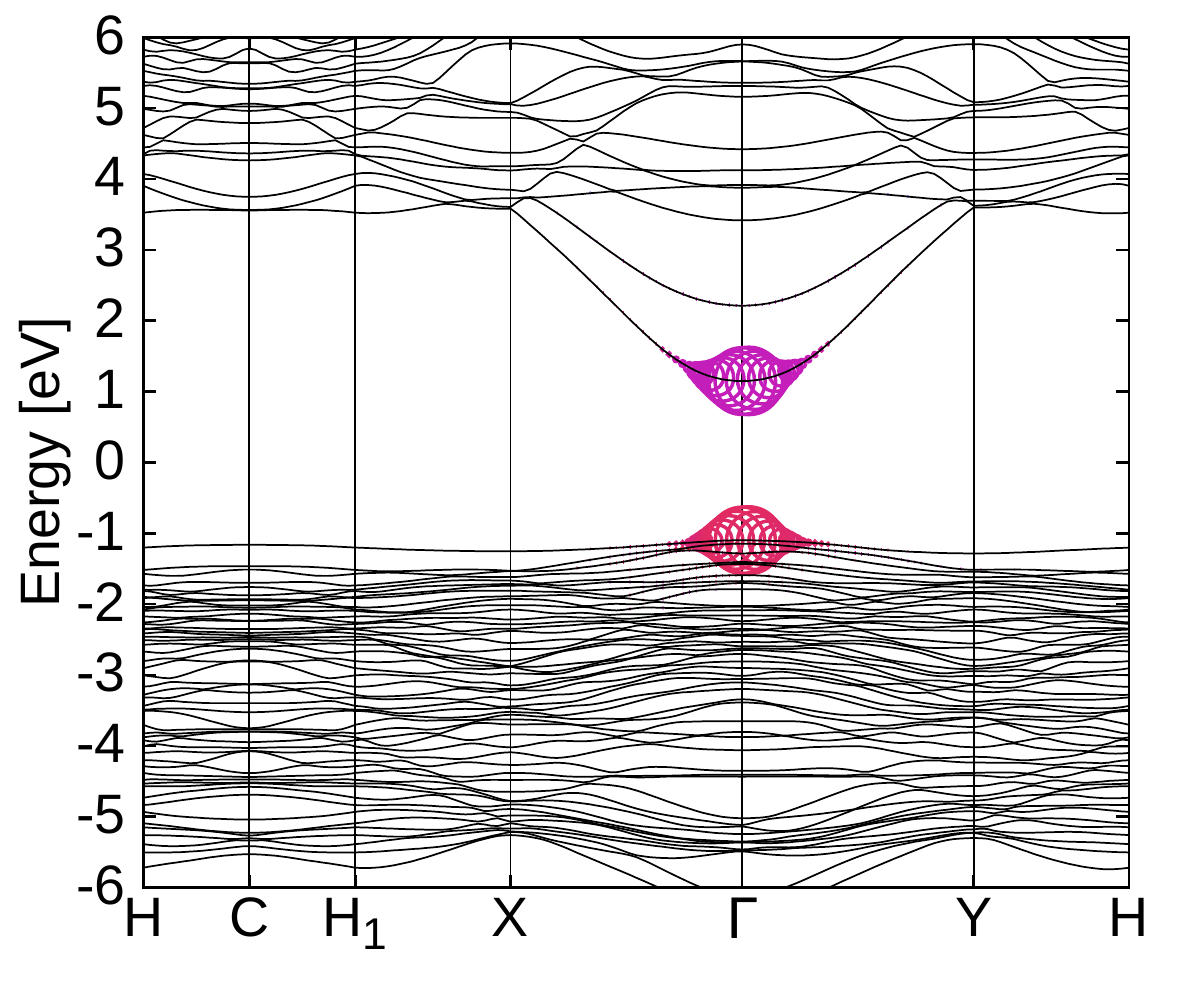}
	\includegraphics[width=0.52\textwidth]{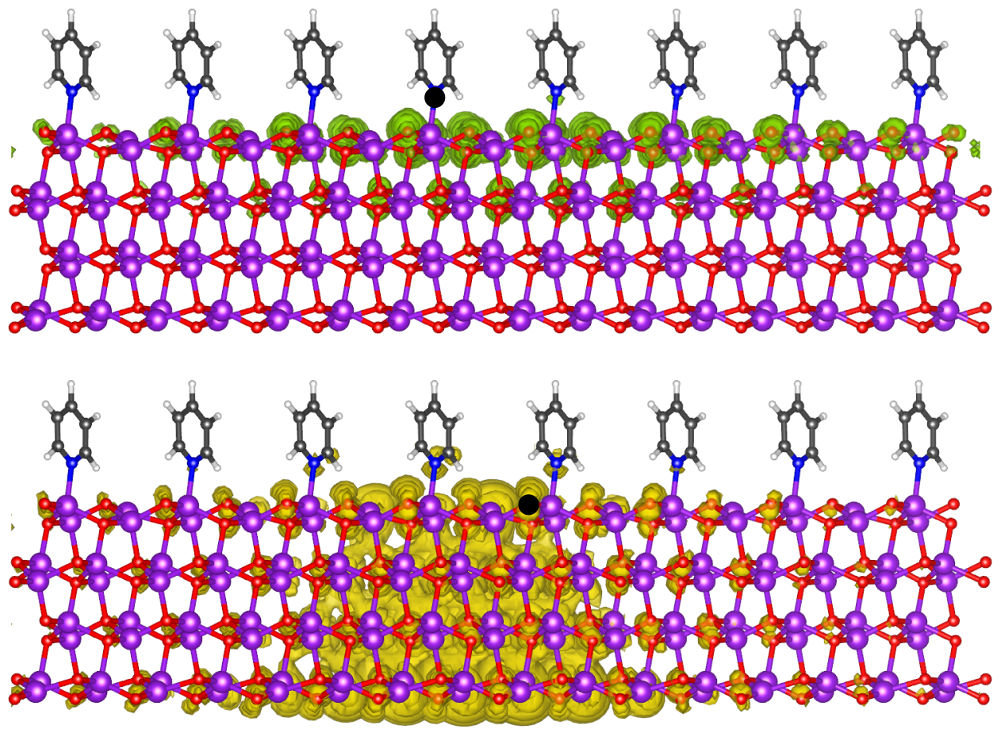}
	\caption{\label{exc_wf1}Reciprocal (left) and real (right) space representation of exciton I depicted in Fig.~\ref{spectrum}. The reciprocal-space representation shows the contributions of individual bands to the exciton as colored circles along a $\ve{k}$ path. The size of the circle is proportional to the transition weight; the colour code indicates the band character going from red (ZnO) over shades of purple (hybridized states) to blue (Py). The real-space representation shows the probability density of finding the hole component of the \textit{e-h} wavefunction given a fixed position of the electron (upper right panel) and vice versa (lower right panel). The electron (hole) probability distribution is depicted in orange (green) with the corresponding hole (electron) position marked by a black circle.}
\end{figure*}

From the solution of the BSE we can gain additional information about the character of the \textit{e-h} pairs in the system. We identify three different types of excitations which we classify as \textit{ZnO-like}, \textit{hybrid}, and \textit{(hybrid) charge-transfer} excitons according to the predominant character of the bands involved in their formation. In the following, we provide a real- and reciprocal-space analysis of those electron-hole pairs which are marked by red bars in the spectrum (Fig.~\ref{spectrum}). We display the excitonic wavefunction, as given by Eq.~\ref{eh_wavefunction}, for fixed positions of the electron and the hole, respectively, in order to inspect their spatial distribution. Additionally, we visualize contributions stemming from different bands, as expressed by Eqs.~\ref{eq:wc} and~\ref{eq:wv} in reciprocal space.

\begin{figure*}
	\includegraphics[width=0.47\textwidth]{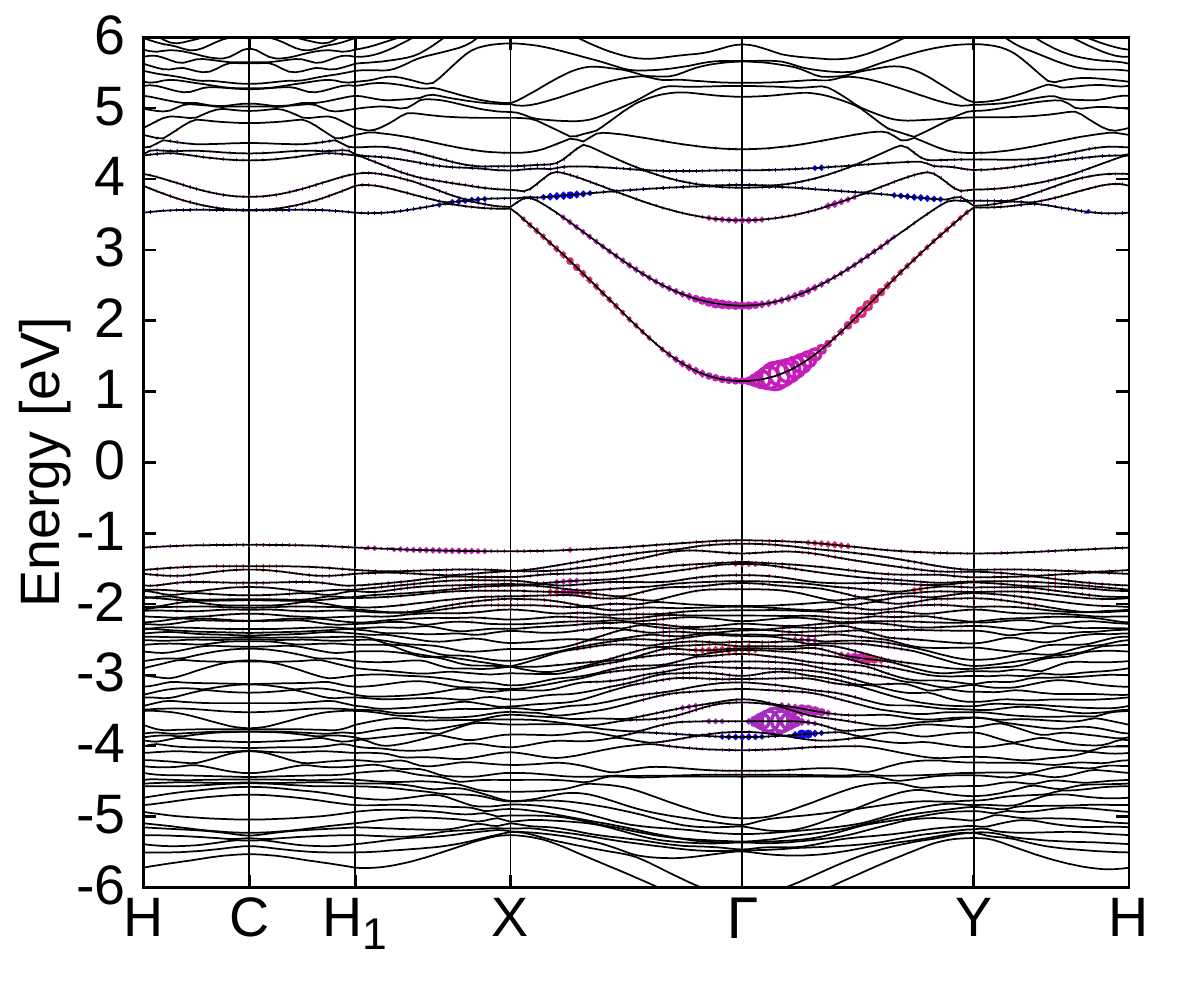}
	\includegraphics[width=0.52\textwidth]{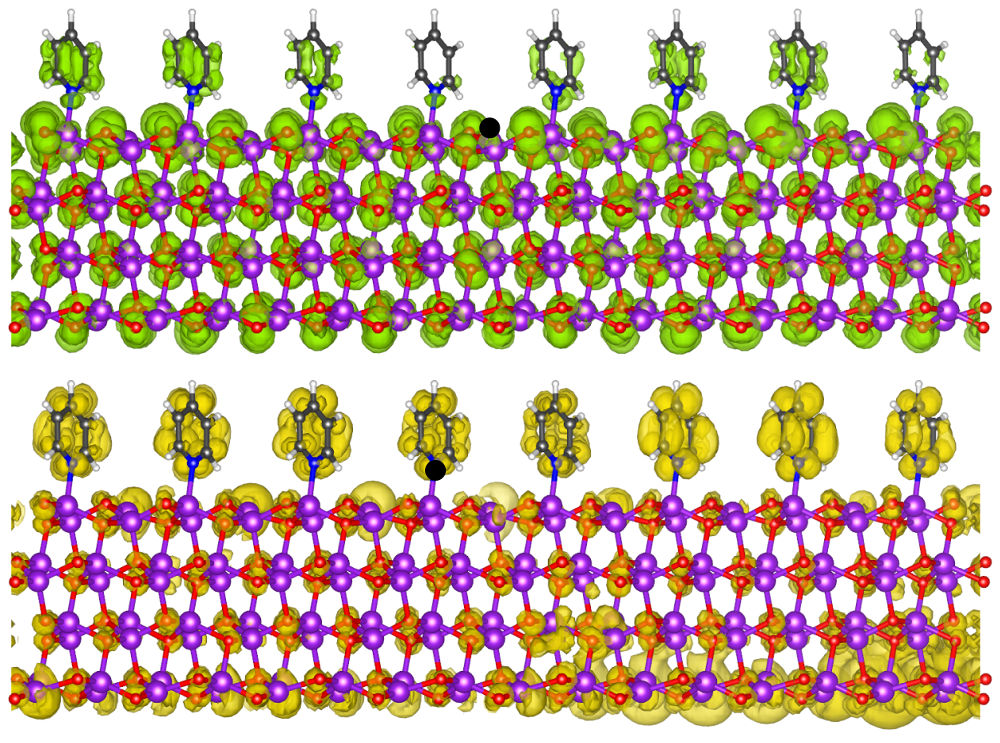}
	\caption{\label{exc_wf2}Same as Fig.~\ref{exc_wf1} for exciton II.}
\end{figure*}

The first exciton (I) at 1.78 eV (Fig.~\ref{spectrum}) is a \textit{hybrid} exciton, stemming from transitions between the hybridized VBM and the CBm at the $\Gamma$ point (Fig.~\ref{exc_wf1}). While the hybrid character of this exciton could already be inferred considering the hybridized nature of the involved electronic states, its spatial extension and shape is provided only in the many-body framework of the BSE. Overall, this exciton is delocalized over the ZnO surface and exhibits nearly centrosymmetric geometry in the xy-plane (see reference system in Fig.~\ref{structure}). The real-space plots in Fig.~\ref{exc_wf1} indicate rather different distributions of the electron and the hole in this \textit{e-h} pair. When the hole is fixed on an oxygen atom of the uppermost ZnO surface layer, the electron distribution is localized in the xy-plane and extended in z direction down across the ZnO layers (Fig.~\ref{exc_wf1}, lower right panel). This feature reflects the bulk character of the CBm depicted in the bottom of Fig.~\ref{bandstructure}. A small amount of the electron distribution also reaches to the nitrogen atom. When the electron is fixed on the N atom of the pyridine bonded to the Zn atom at the surface, the hole distribution is limited to the uppermost ZnO layer, with minor contributions from the second layer. This analysis shows that the first exciton has predominantly ZnO character bearing therefore a weak hybrid nature. Also the manifold of excitations between exciton I and exciton II in Fig.~\ref{spectrum} exhibit similar character.

Excitons with a pronounced hybrid nature, albeit of significantly low intensity compared to the first one, can be found at higher energies. Representative of this type is the exciton at 4.74~eV (Fig.~\ref{exc_wf2}), labeled II in Fig.~\ref{spectrum}. Its main character stems from transitions along the $\Gamma$-Y path near the $\Gamma$ point, from VBM-33 to the lowest conduction band, the VBM-33 being a deep lying hybrid band in the valence region that has oxygen, nitrogen, and carbon contributions (Fig.~\ref{bandstructure}). With the electron fixed on a surface-layer oxygen atom, the hole distribution appears entirely delocalized over both pyridine and ZnO, demonstrating its hybrid character. When fixing the hole on the nitrogen, the hybrid nature of the electron distribution becomes evident. Here we observe a significant weight of the electron-hole pair distribution on the lowest ZnO layer which indicates a confinement effect. We conclude that the delocalization and, related to this, the electron-hole binding strength, may depend on the thickness of the ZnO substrate. More specific, the character of such electron-hole pair may become even more ZnO-dominated with increasing ZnO thickness.

\begin{figure*}
	\includegraphics[width=0.47\textwidth]{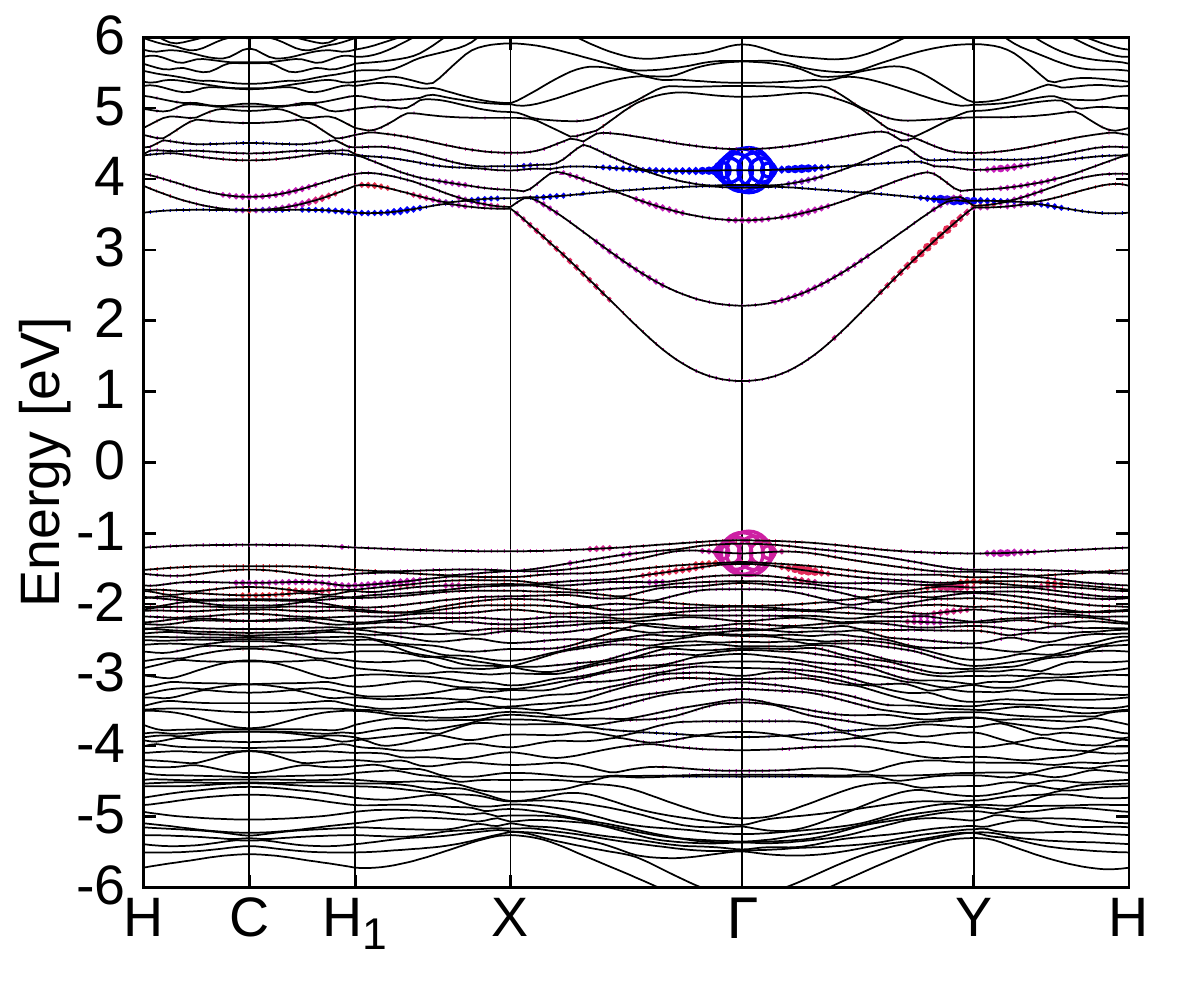}
	\includegraphics[width=0.52\textwidth]{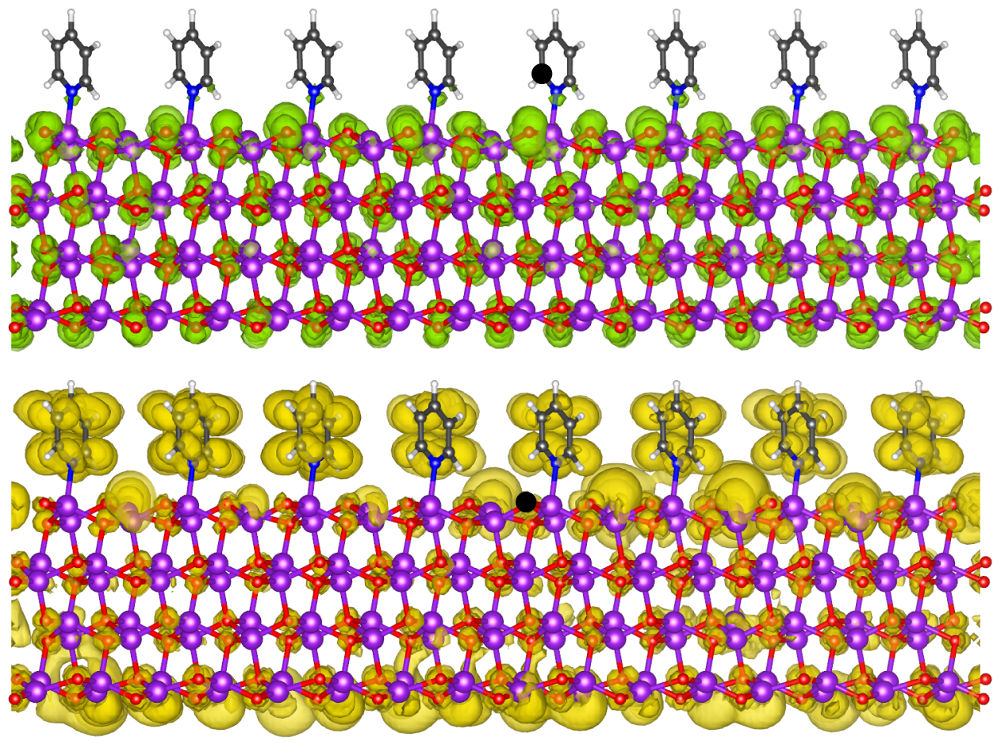}
	\caption{\label{exc_wf3}Same as Fig.~\ref{exc_wf1} for exciton III.}
\end{figure*}

Another type of exciton, that can be found in the UV region, has \textit{charge-transfer} nature, being characterized by spatial separation of the electron and the hole across the interface. When the electron is situated on the organic side, the hole resides mainly on the inorganic component, and vice versa. Such excitons involve transitions form pure pyridine bands to ZnO-dominated bands. This scenario is realized in the exciton at 5.31~eV, composed of $\Gamma$-point transitions from VBM-2, which displays ZnO character, to Py(LUMO+1). By plotting the excitonic wavefunction in real space (Fig.~\ref{exc_wf3}) we confirm the charge-transfer character of this excitation. Fixing the electron position on a carbon atom, the hole distribution is spread over ZnO (upper right panel), while fixing the hole on an oxygen atom (lower right panel) reveals an electron distribution located mainly on pyridine. In the latter case, we additionally observe some intensity on ZnO. This indicates that there are mixed-in contributions from hybrid or ZnO states along a different $\ve{k}$-path.

\begin{figure*}
	\includegraphics[width=0.47\textwidth]{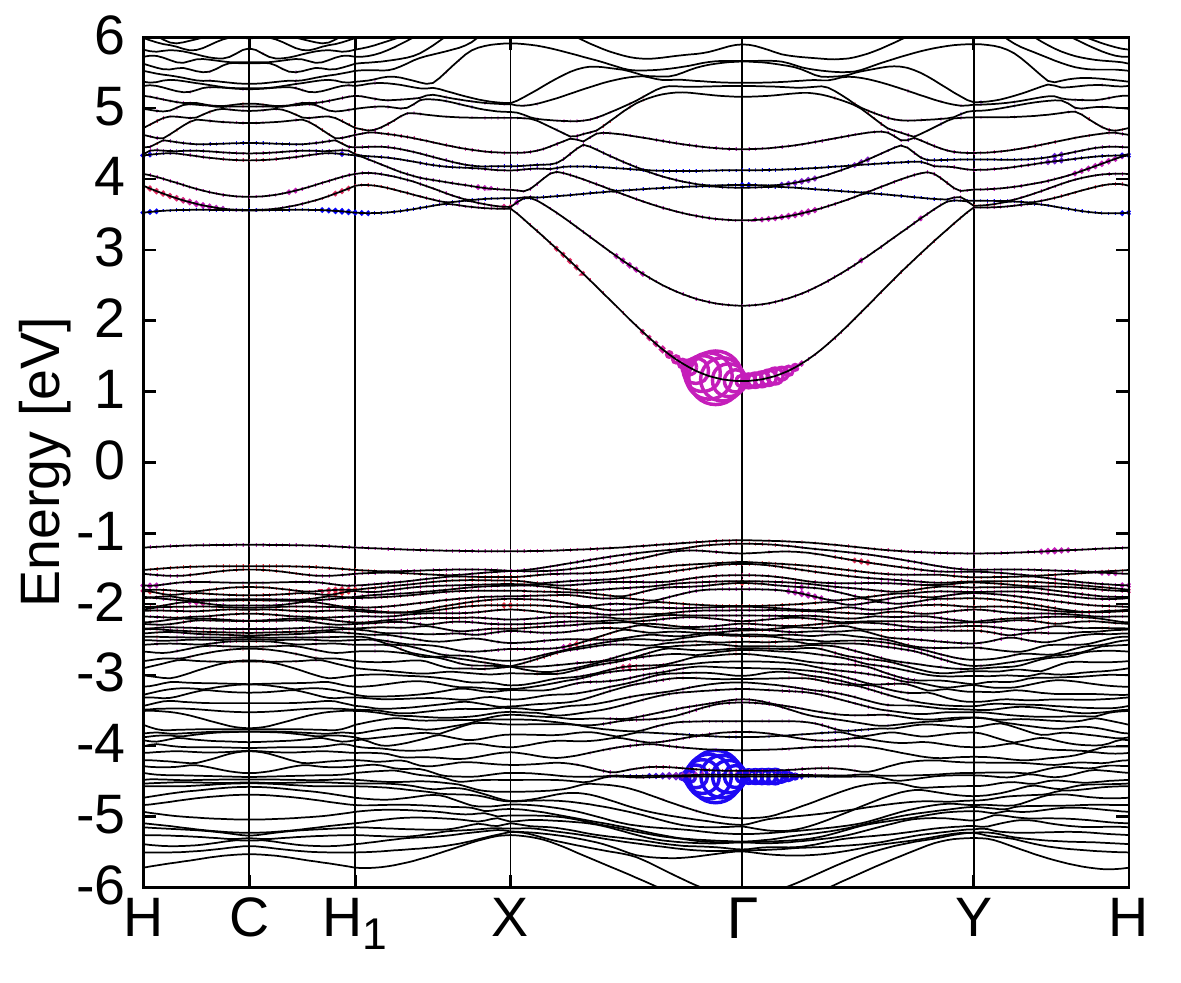}
	\includegraphics[width=0.52\textwidth]{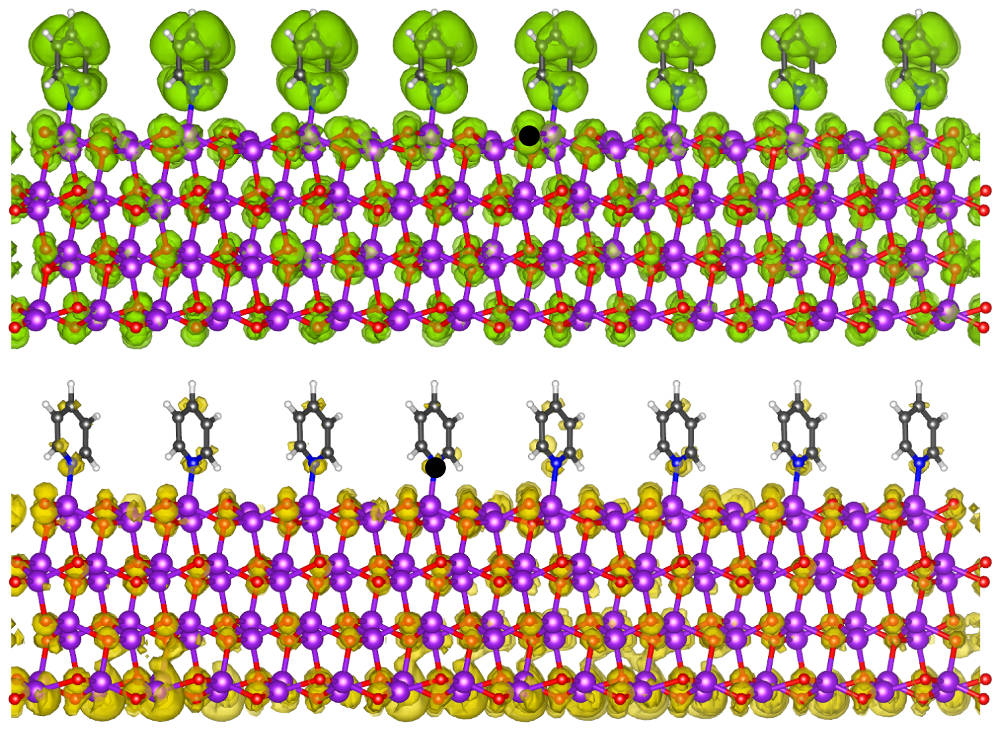}
	\caption{\label{exc_wf4}Same as Fig.~\ref{exc_wf1} for exciton IV.}
\end{figure*}

Also exciton IV at 5.49~eV exhibits charge-transfer nature (see Fig.~\ref{exc_wf4}). However, here, the involved pyridine band lies in the valence region, as reflected in the hole distribution. The reciprocal-space analysis shows transitions from Py(HOMO-2) to the CBm around the $\Gamma$ point, with the transition weights being larger along X-$\Gamma$ than along $\Gamma$-Y direction. Fixing the electron position on the oxygen atom exhibits a delocalized Py(HOMO-2)-like hole distribution on pyridine. Fixing the hole position on nitrogen, we observe the electron distribution to reside mainly on ZnO.

This wealth of excitations with charge-transfer character appears in the UV region of the spectrum, well above its onset at visible frequencies. In essence, our results indicate that the chemisorption of the molecule on the ZnO surface tends to largely preserve the optical absorption features of the inorganic material in the onset region. This is related to the small size of pyridine and thus its sizeable band gap, which is a few eV larger than the one of the ZnO surface. This difference does not represent the optimal scenario for opto-electronic applications where the gaps of the constituents would ideally be of similar size. This criterion can be met by choosing molecules with a more extended backbone. For instance, attaching more phenyl rings to the pyridine molecule would shift the HOMO and LUMO of the organic components to lower energies, and thus closer to the frontier region of the ZnO surface. 

\section{Conclusions} 
In summary, we have shown how many-body effects govern the opto-electronic excitations of prototypical inorganic/organic hybrid interface, formed by a monolayer of pyridine molecules chemisorbed on the non-polar ZnO($10\overline{1}0$) surface. This system exhibits a type-I band alignment, with hybridized frontier states. We emphasize that the DFT band structure of such systems, based on semi-local xc functionals, is meaningless in view of the distinctively different self-energy corrections to different band states. The $G_0W_0$@LDA approach is not fully quantitative due to the perturbative nature of the $G_0W_0$ approximation and, consequently, its starting-point dependence. Still, it provides an important step forward to handle systems of this complexity. We have computed the optical excitation spectrum by solving the BSE and analyzed the character of various optical excitations in terms of the character of contributing states and the real-space extension of the electron-hole wavefunction. More specific, with four different examples we have shown that hybrid excitons as well as (hybrid) charge-transfer excitons can occur in such materials which appear in the UV region. We observe charge-transfer excitons with the electron distributed on the organic constituent and the hole on the inorganic one as well as excitons with a reverse electron-hole distribution. At the onset of the spectrum, we find strongly bound bright excitons, the lowest one being a ZnO-dominated hybrid exciton with a binding energy of 0.4~eV. 

Our results indicate that the complexity of the optical excitations in such hybrid interfaces is ruled by a tight interplay between quantum confinement, level alignment, and electronic hybridization. In the specific interface investigated here, the distinctly different band gaps of the two components make the ZnO surface dominating  the low-energy range of the spectrum. Excitations with pronounced hybrid and charge-transfer character appear only well above the absorption onset in the UV region. However, upon appropriate choice of the molecular component, they could be shifted to lower energies to meet the requirements for a hybrid interface for opto-electronics.

\begin{acknowledgments}
Work funded by the German Research Foundation (DFG), through the Collaborative Research Center 951. We are grateful to Sebastian Tillack for providing the Wannier interpolation and Benjamin Aurich for speeding up the BSE calculation. Both are kindly acknowledged for fruitful discussions.
\end{acknowledgments}


\begin{thebibliography}{46}%
\makeatletter
\providecommand \@ifxundefined [1]{%
 \@ifx{#1\undefined}
}%
\providecommand \@ifnum [1]{%
 \ifnum #1\expandafter \@firstoftwo
 \else \expandafter \@secondoftwo
 \fi
}%
\providecommand \@ifx [1]{%
 \ifx #1\expandafter \@firstoftwo
 \else \expandafter \@secondoftwo
 \fi
}%
\providecommand \natexlab [1]{#1}%
\providecommand \enquote  [1]{``#1''}%
\providecommand \bibnamefont  [1]{#1}%
\providecommand \bibfnamefont [1]{#1}%
\providecommand \citenamefont [1]{#1}%
\providecommand \href@noop [0]{\@secondoftwo}%
\providecommand \href [0]{\begingroup \@sanitize@url \@href}%
\providecommand \@href[1]{\@@startlink{#1}\@@href}%
\providecommand \@@href[1]{\endgroup#1\@@endlink}%
\providecommand \@sanitize@url [0]{\catcode `\\12\catcode `\$12\catcode
  `\&12\catcode `\#12\catcode `\^12\catcode `\_12\catcode `\%12\relax}%
\providecommand \@@startlink[1]{}%
\providecommand \@@endlink[0]{}%
\providecommand \url  [0]{\begingroup\@sanitize@url \@url }%
\providecommand \@url [1]{\endgroup\@href {#1}{\urlprefix }}%
\providecommand \urlprefix  [0]{URL }%
\providecommand \Eprint [0]{\href }%
\providecommand \doibase [0]{http://dx.doi.org/}%
\providecommand \selectlanguage [0]{\@gobble}%
\providecommand \bibinfo  [0]{\@secondoftwo}%
\providecommand \bibfield  [0]{\@secondoftwo}%
\providecommand \translation [1]{[#1]}%
\providecommand \BibitemOpen [0]{}%
\providecommand \bibitemStop [0]{}%
\providecommand \bibitemNoStop [0]{.\EOS\space}%
\providecommand \EOS [0]{\spacefactor3000\relax}%
\providecommand \BibitemShut  [1]{\csname bibitem#1\endcsname}%
\let\auto@bib@innerbib\@empty
\bibitem [{\citenamefont {Ishii}\ \emph {et~al.}(1999)\citenamefont {Ishii},
  \citenamefont {Sugiyama}, \citenamefont {Ito},\ and\ \citenamefont
  {Seki}}]{ishii1999energy}%
  \BibitemOpen
  \bibfield  {author} {\bibinfo {author} {\bibfnamefont {H.}~\bibnamefont
  {Ishii}}, \bibinfo {author} {\bibfnamefont {K.}~\bibnamefont {Sugiyama}},
  \bibinfo {author} {\bibfnamefont {E.}~\bibnamefont {Ito}}, \ and\ \bibinfo
  {author} {\bibfnamefont {K.}~\bibnamefont {Seki}},\ }\href@noop {} {\bibfield
   {journal} {\bibinfo  {journal} {Adv. Mater.}\ }\textbf {\bibinfo {volume}
  {11}},\ \bibinfo {pages} {605} (\bibinfo {year} {1999})}\BibitemShut
  {NoStop}%
\bibitem [{\citenamefont {Agranovich}\ \emph {et~al.}(2011)\citenamefont
  {Agranovich}, \citenamefont {Gartstein},\ and\ \citenamefont
  {Litinskaya}}]{Hybrid_review}%
  \BibitemOpen
  \bibfield  {author} {\bibinfo {author} {\bibfnamefont {V.}~\bibnamefont
  {Agranovich}}, \bibinfo {author} {\bibfnamefont {Y.~N.}\ \bibnamefont
  {Gartstein}}, \ and\ \bibinfo {author} {\bibfnamefont {M.}~\bibnamefont
  {Litinskaya}},\ }\href@noop {} {\bibfield  {journal} {\bibinfo  {journal}
  {Chem. Rev.}\ }\textbf {\bibinfo {volume} {111}},\ \bibinfo {pages} {5179}
  (\bibinfo {year} {2011})}\BibitemShut {NoStop}%
\bibitem [{\citenamefont {Wright}\ and\ \citenamefont
  {Uddin}(2012)}]{wright2012organic}%
  \BibitemOpen
  \bibfield  {author} {\bibinfo {author} {\bibfnamefont {M.}~\bibnamefont
  {Wright}}\ and\ \bibinfo {author} {\bibfnamefont {A.}~\bibnamefont {Uddin}},\
  }\href@noop {} {\bibfield  {journal} {\bibinfo  {journal} {Sol. Energy Mater
  Sol. Cells}\ }\textbf {\bibinfo {volume} {107}},\ \bibinfo {pages} {87}
  (\bibinfo {year} {2012})}\BibitemShut {NoStop}%
\bibitem [{\citenamefont {Coe}\ \emph {et~al.}(2002)\citenamefont {Coe},
  \citenamefont {Woo}, \citenamefont {Bawendi},\ and\ \citenamefont
  {Bulovi{\'c}}}]{Coe}%
  \BibitemOpen
  \bibfield  {author} {\bibinfo {author} {\bibfnamefont {S.}~\bibnamefont
  {Coe}}, \bibinfo {author} {\bibfnamefont {W.-K.}\ \bibnamefont {Woo}},
  \bibinfo {author} {\bibfnamefont {M.}~\bibnamefont {Bawendi}}, \ and\
  \bibinfo {author} {\bibfnamefont {V.}~\bibnamefont {Bulovi{\'c}}},\
  }\href@noop {} {\bibfield  {journal} {\bibinfo  {journal} {Nature}\ }\textbf
  {\bibinfo {volume} {420}},\ \bibinfo {pages} {800} (\bibinfo {year}
  {2002})}\BibitemShut {NoStop}%
\bibitem [{\citenamefont {Xu}\ and\ \citenamefont
  {Qiao}(2011)}]{xu2011conjugated}%
  \BibitemOpen
  \bibfield  {author} {\bibinfo {author} {\bibfnamefont {T.}~\bibnamefont
  {Xu}}\ and\ \bibinfo {author} {\bibfnamefont {Q.}~\bibnamefont {Qiao}},\
  }\href@noop {} {\bibfield  {journal} {\bibinfo  {journal} {Energy Environ.
  Sci.}\ }\textbf {\bibinfo {volume} {4}},\ \bibinfo {pages} {2700} (\bibinfo
  {year} {2011})}\BibitemShut {NoStop}%
\bibitem [{\citenamefont {Chan}\ \emph {et~al.}(2012)\citenamefont {Chan},
  \citenamefont {Chen}, \citenamefont {Lin},\ and\ \citenamefont
  {Chen}}]{chan2012direct}%
  \BibitemOpen
  \bibfield  {author} {\bibinfo {author} {\bibfnamefont {M.}~\bibnamefont
  {Chan}}, \bibinfo {author} {\bibfnamefont {J.}~\bibnamefont {Chen}}, \bibinfo
  {author} {\bibfnamefont {T.}~\bibnamefont {Lin}}, \ and\ \bibinfo {author}
  {\bibfnamefont {Y.}~\bibnamefont {Chen}},\ }\href@noop {} {\bibfield
  {journal} {\bibinfo  {journal} {Appl. Phys. Lett.}\ }\textbf {\bibinfo
  {volume} {100}},\ \bibinfo {pages} {021912} (\bibinfo {year}
  {2012})}\BibitemShut {NoStop}%
\bibitem [{\citenamefont {Nagata}\ \emph {et~al.}(2013)\citenamefont {Nagata},
  \citenamefont {Oh}, \citenamefont {Yamashita}, \citenamefont {Yoshikawa},
  \citenamefont {Ikeno}, \citenamefont {Kobayashi}, \citenamefont {Chikyow},\
  and\ \citenamefont {Wakayama}}]{nagata2013photoelectron}%
  \BibitemOpen
  \bibfield  {author} {\bibinfo {author} {\bibfnamefont {T.}~\bibnamefont
  {Nagata}}, \bibinfo {author} {\bibfnamefont {S.}~\bibnamefont {Oh}}, \bibinfo
  {author} {\bibfnamefont {Y.}~\bibnamefont {Yamashita}}, \bibinfo {author}
  {\bibfnamefont {H.}~\bibnamefont {Yoshikawa}}, \bibinfo {author}
  {\bibfnamefont {N.}~\bibnamefont {Ikeno}}, \bibinfo {author} {\bibfnamefont
  {K.}~\bibnamefont {Kobayashi}}, \bibinfo {author} {\bibfnamefont
  {T.}~\bibnamefont {Chikyow}}, \ and\ \bibinfo {author} {\bibfnamefont
  {Y.}~\bibnamefont {Wakayama}},\ }\href@noop {} {\bibfield  {journal}
  {\bibinfo  {journal} {Appl. Phys. Lett.}\ }\textbf {\bibinfo {volume}
  {102}},\ \bibinfo {pages} {17} (\bibinfo {year} {2013})}\BibitemShut
  {NoStop}%
\bibitem [{\citenamefont {Gao}\ \emph {et~al.}(2013)\citenamefont {Gao},
  \citenamefont {Ren},\ and\ \citenamefont {Wang}}]{gao2013renaissance}%
  \BibitemOpen
  \bibfield  {author} {\bibinfo {author} {\bibfnamefont {F.}~\bibnamefont
  {Gao}}, \bibinfo {author} {\bibfnamefont {S.}~\bibnamefont {Ren}}, \ and\
  \bibinfo {author} {\bibfnamefont {J.}~\bibnamefont {Wang}},\ }\href@noop {}
  {\bibfield  {journal} {\bibinfo  {journal} {Energy Environ. Sci.}\ }\textbf
  {\bibinfo {volume} {6}},\ \bibinfo {pages} {2020} (\bibinfo {year}
  {2013})}\BibitemShut {NoStop}%
\bibitem [{\citenamefont {Bianchi}\ \emph {et~al.}(2014)\citenamefont
  {Bianchi}, \citenamefont {Sadofev}, \citenamefont {Schlesinger},
  \citenamefont {Kobin}, \citenamefont {Hecht}, \citenamefont {Koch},
  \citenamefont {Henneberger},\ and\ \citenamefont
  {Blumstengel}}]{bianchi2014cascade}%
  \BibitemOpen
  \bibfield  {author} {\bibinfo {author} {\bibfnamefont {F.}~\bibnamefont
  {Bianchi}}, \bibinfo {author} {\bibfnamefont {S.}~\bibnamefont {Sadofev}},
  \bibinfo {author} {\bibfnamefont {R.}~\bibnamefont {Schlesinger}}, \bibinfo
  {author} {\bibfnamefont {B.}~\bibnamefont {Kobin}}, \bibinfo {author}
  {\bibfnamefont {S.}~\bibnamefont {Hecht}}, \bibinfo {author} {\bibfnamefont
  {N.}~\bibnamefont {Koch}}, \bibinfo {author} {\bibfnamefont {F.}~\bibnamefont
  {Henneberger}}, \ and\ \bibinfo {author} {\bibfnamefont {S.}~\bibnamefont
  {Blumstengel}},\ }\href@noop {} {\bibfield  {journal} {\bibinfo  {journal}
  {Appl. Phys. Lett.}\ }\textbf {\bibinfo {volume} {105}},\ \bibinfo {pages}
  {183\_1} (\bibinfo {year} {2014})}\BibitemShut {NoStop}%
\bibitem [{\citenamefont {Piersimoni}\ \emph {et~al.}(2015)\citenamefont
  {Piersimoni}, \citenamefont {Schlesinger}, \citenamefont {Benduhn},
  \citenamefont {Spoltore}, \citenamefont {Reiter}, \citenamefont {Lange},
  \citenamefont {Koch}, \citenamefont {Vandewal},\ and\ \citenamefont
  {Neher}}]{piersimoni2015charge}%
  \BibitemOpen
  \bibfield  {author} {\bibinfo {author} {\bibfnamefont {F.}~\bibnamefont
  {Piersimoni}}, \bibinfo {author} {\bibfnamefont {R.}~\bibnamefont
  {Schlesinger}}, \bibinfo {author} {\bibfnamefont {J.}~\bibnamefont
  {Benduhn}}, \bibinfo {author} {\bibfnamefont {D.}~\bibnamefont {Spoltore}},
  \bibinfo {author} {\bibfnamefont {S.}~\bibnamefont {Reiter}}, \bibinfo
  {author} {\bibfnamefont {I.}~\bibnamefont {Lange}}, \bibinfo {author}
  {\bibfnamefont {N.}~\bibnamefont {Koch}}, \bibinfo {author} {\bibfnamefont
  {K.}~\bibnamefont {Vandewal}}, \ and\ \bibinfo {author} {\bibfnamefont
  {D.}~\bibnamefont {Neher}},\ }\href@noop {} {\bibfield  {journal} {\bibinfo
  {journal} {J. Phys. Chem. Lett.}\ }\textbf {\bibinfo {volume} {6}},\ \bibinfo
  {pages} {500} (\bibinfo {year} {2015})}\BibitemShut {NoStop}%
\bibitem [{\citenamefont {Blumstengel}\ \emph {et~al.}(2010)\citenamefont
  {Blumstengel}, \citenamefont {Glowatzki}, \citenamefont {Sadofev},
  \citenamefont {Koch}, \citenamefont {Kowarik}, \citenamefont {Rabe},\ and\
  \citenamefont {Henneberger}}]{blumstengel2010band}%
  \BibitemOpen
  \bibfield  {author} {\bibinfo {author} {\bibfnamefont {S.}~\bibnamefont
  {Blumstengel}}, \bibinfo {author} {\bibfnamefont {H.}~\bibnamefont
  {Glowatzki}}, \bibinfo {author} {\bibfnamefont {S.}~\bibnamefont {Sadofev}},
  \bibinfo {author} {\bibfnamefont {N.}~\bibnamefont {Koch}}, \bibinfo {author}
  {\bibfnamefont {S.}~\bibnamefont {Kowarik}}, \bibinfo {author} {\bibfnamefont
  {J.~P.}\ \bibnamefont {Rabe}}, \ and\ \bibinfo {author} {\bibfnamefont
  {F.}~\bibnamefont {Henneberger}},\ }\href@noop {} {\bibfield  {journal}
  {\bibinfo  {journal} {Phys. Chem. Chem. Phys.}\ }\textbf {\bibinfo {volume}
  {12}},\ \bibinfo {pages} {11642} (\bibinfo {year} {2010})}\BibitemShut
  {NoStop}%
\bibitem [{\citenamefont {Wood}\ \emph {et~al.}(2012)\citenamefont {Wood},
  \citenamefont {Li}, \citenamefont {Winget},\ and\ \citenamefont
  {Bredas}}]{wood2012binding}%
  \BibitemOpen
  \bibfield  {author} {\bibinfo {author} {\bibfnamefont {C.}~\bibnamefont
  {Wood}}, \bibinfo {author} {\bibfnamefont {H.}~\bibnamefont {Li}}, \bibinfo
  {author} {\bibfnamefont {P.}~\bibnamefont {Winget}}, \ and\ \bibinfo {author}
  {\bibfnamefont {J.-L.}\ \bibnamefont {Bredas}},\ }\href@noop {} {\bibfield
  {journal} {\bibinfo  {journal} {J. Phys. Chem. C}\ }\textbf {\bibinfo
  {volume} {116}},\ \bibinfo {pages} {19125} (\bibinfo {year}
  {2012})}\BibitemShut {NoStop}%
\bibitem [{\citenamefont {Xu}\ \emph {et~al.}(2013)\citenamefont {Xu},
  \citenamefont {Hofmann}, \citenamefont {Schlesinger}, \citenamefont
  {Winkler}, \citenamefont {Frisch}, \citenamefont {Niederhausen},
  \citenamefont {Vollmer}, \citenamefont {Blumstengel}, \citenamefont
  {Henneberger}, \citenamefont {Koch}, \citenamefont {Rinke},\ and\
  \citenamefont {Scheffler}}]{xu2013}%
  \BibitemOpen
  \bibfield  {author} {\bibinfo {author} {\bibfnamefont {Y.}~\bibnamefont
  {Xu}}, \bibinfo {author} {\bibfnamefont {O.~T.}\ \bibnamefont {Hofmann}},
  \bibinfo {author} {\bibfnamefont {R.}~\bibnamefont {Schlesinger}}, \bibinfo
  {author} {\bibfnamefont {S.}~\bibnamefont {Winkler}}, \bibinfo {author}
  {\bibfnamefont {J.}~\bibnamefont {Frisch}}, \bibinfo {author} {\bibfnamefont
  {J.}~\bibnamefont {Niederhausen}}, \bibinfo {author} {\bibfnamefont
  {A.}~\bibnamefont {Vollmer}}, \bibinfo {author} {\bibfnamefont
  {S.}~\bibnamefont {Blumstengel}}, \bibinfo {author} {\bibfnamefont
  {F.}~\bibnamefont {Henneberger}}, \bibinfo {author} {\bibfnamefont
  {N.}~\bibnamefont {Koch}}, \bibinfo {author} {\bibfnamefont {P.}~\bibnamefont
  {Rinke}}, \ and\ \bibinfo {author} {\bibfnamefont {M.}~\bibnamefont
  {Scheffler}},\ }\href {\doibase 10.1103/PhysRevLett.111.226802} {\bibfield
  {journal} {\bibinfo  {journal} {Phys. Rev. Lett.}\ }\textbf {\bibinfo
  {volume} {111}},\ \bibinfo {pages} {226802} (\bibinfo {year}
  {2013})}\BibitemShut {NoStop}%
\bibitem [{\citenamefont {Schlesinger}\ \emph {et~al.}(2013)\citenamefont
  {Schlesinger}, \citenamefont {Xu}, \citenamefont {Hofmann}, \citenamefont
  {Winkler}, \citenamefont {Frisch}, \citenamefont {Niederhausen},
  \citenamefont {Vollmer}, \citenamefont {Blumstengel}, \citenamefont
  {Henneberger}, \citenamefont {Rinke}, \citenamefont {Scheffler},\ and\
  \citenamefont {Koch}}]{Schlesinger2013}%
  \BibitemOpen
  \bibfield  {author} {\bibinfo {author} {\bibfnamefont {R.}~\bibnamefont
  {Schlesinger}}, \bibinfo {author} {\bibfnamefont {Y.}~\bibnamefont {Xu}},
  \bibinfo {author} {\bibfnamefont {O.~T.}\ \bibnamefont {Hofmann}}, \bibinfo
  {author} {\bibfnamefont {S.}~\bibnamefont {Winkler}}, \bibinfo {author}
  {\bibfnamefont {J.}~\bibnamefont {Frisch}}, \bibinfo {author} {\bibfnamefont
  {J.}~\bibnamefont {Niederhausen}}, \bibinfo {author} {\bibfnamefont
  {A.}~\bibnamefont {Vollmer}}, \bibinfo {author} {\bibfnamefont
  {S.}~\bibnamefont {Blumstengel}}, \bibinfo {author} {\bibfnamefont
  {F.}~\bibnamefont {Henneberger}}, \bibinfo {author} {\bibfnamefont
  {P.}~\bibnamefont {Rinke}}, \bibinfo {author} {\bibfnamefont
  {M.}~\bibnamefont {Scheffler}}, \ and\ \bibinfo {author} {\bibfnamefont
  {N.}~\bibnamefont {Koch}},\ }\href {\doibase 10.1103/PhysRevB.87.155311}
  {\bibfield  {journal} {\bibinfo  {journal} {Phys. Rev. B}\ }\textbf {\bibinfo
  {volume} {87}},\ \bibinfo {pages} {155311} (\bibinfo {year}
  {2013})}\BibitemShut {NoStop}%
\bibitem [{\citenamefont {Musselman}\ \emph {et~al.}(2014)\citenamefont
  {Musselman}, \citenamefont {Albert-Seifried}, \citenamefont {Hoye},
  \citenamefont {Sadhanala}, \citenamefont {Mu{\~n}oz-Rojas}, \citenamefont
  {MacManus-Driscoll},\ and\ \citenamefont {Friend}}]{musselman2014improved}%
  \BibitemOpen
  \bibfield  {author} {\bibinfo {author} {\bibfnamefont {K.~P.}\ \bibnamefont
  {Musselman}}, \bibinfo {author} {\bibfnamefont {S.}~\bibnamefont
  {Albert-Seifried}}, \bibinfo {author} {\bibfnamefont {R.~L.}\ \bibnamefont
  {Hoye}}, \bibinfo {author} {\bibfnamefont {A.}~\bibnamefont {Sadhanala}},
  \bibinfo {author} {\bibfnamefont {D.}~\bibnamefont {Mu{\~n}oz-Rojas}},
  \bibinfo {author} {\bibfnamefont {J.~L.}\ \bibnamefont {MacManus-Driscoll}},
  \ and\ \bibinfo {author} {\bibfnamefont {R.~H.}\ \bibnamefont {Friend}},\
  }\href@noop {} {\bibfield  {journal} {\bibinfo  {journal} {Adv. Funct.
  Mater.}\ }\textbf {\bibinfo {volume} {24}},\ \bibinfo {pages} {3562}
  (\bibinfo {year} {2014})}\BibitemShut {NoStop}%
\bibitem [{\citenamefont {Lange}\ \emph {et~al.}(2014)\citenamefont {Lange},
  \citenamefont {Reiter}, \citenamefont {P{\"a}tzel}, \citenamefont {Zykov},
  \citenamefont {Nefedov}, \citenamefont {Hildebrandt}, \citenamefont {Hecht},
  \citenamefont {Kowarik}, \citenamefont {W{\"o}ll}, \citenamefont {Heimel}
  \emph {et~al.}}]{lange2014tuning}%
  \BibitemOpen
  \bibfield  {author} {\bibinfo {author} {\bibfnamefont {I.}~\bibnamefont
  {Lange}}, \bibinfo {author} {\bibfnamefont {S.}~\bibnamefont {Reiter}},
  \bibinfo {author} {\bibfnamefont {M.}~\bibnamefont {P{\"a}tzel}}, \bibinfo
  {author} {\bibfnamefont {A.}~\bibnamefont {Zykov}}, \bibinfo {author}
  {\bibfnamefont {A.}~\bibnamefont {Nefedov}}, \bibinfo {author} {\bibfnamefont
  {J.}~\bibnamefont {Hildebrandt}}, \bibinfo {author} {\bibfnamefont
  {S.}~\bibnamefont {Hecht}}, \bibinfo {author} {\bibfnamefont
  {S.}~\bibnamefont {Kowarik}}, \bibinfo {author} {\bibfnamefont
  {C.}~\bibnamefont {W{\"o}ll}}, \bibinfo {author} {\bibfnamefont
  {G.}~\bibnamefont {Heimel}},  \emph {et~al.},\ }\href@noop {} {\bibfield
  {journal} {\bibinfo  {journal} {Adv. Funct. Mater.}\ }\textbf {\bibinfo
  {volume} {24}},\ \bibinfo {pages} {7014} (\bibinfo {year}
  {2014})}\BibitemShut {NoStop}%
\bibitem [{\citenamefont {Timpel}\ \emph {et~al.}(2014)\citenamefont {Timpel},
  \citenamefont {Nardi}, \citenamefont {Krause}, \citenamefont {Ligorio},
  \citenamefont {Christodoulou}, \citenamefont {Pasquali}, \citenamefont
  {Giglia}, \citenamefont {Frisch}, \citenamefont {Wegner}, \citenamefont
  {Moras} \emph {et~al.}}]{timpel2014surface}%
  \BibitemOpen
  \bibfield  {author} {\bibinfo {author} {\bibfnamefont {M.}~\bibnamefont
  {Timpel}}, \bibinfo {author} {\bibfnamefont {M.~V.}\ \bibnamefont {Nardi}},
  \bibinfo {author} {\bibfnamefont {S.}~\bibnamefont {Krause}}, \bibinfo
  {author} {\bibfnamefont {G.}~\bibnamefont {Ligorio}}, \bibinfo {author}
  {\bibfnamefont {C.}~\bibnamefont {Christodoulou}}, \bibinfo {author}
  {\bibfnamefont {L.}~\bibnamefont {Pasquali}}, \bibinfo {author}
  {\bibfnamefont {A.}~\bibnamefont {Giglia}}, \bibinfo {author} {\bibfnamefont
  {J.}~\bibnamefont {Frisch}}, \bibinfo {author} {\bibfnamefont
  {B.}~\bibnamefont {Wegner}}, \bibinfo {author} {\bibfnamefont
  {P.}~\bibnamefont {Moras}},  \emph {et~al.},\ }\href@noop {} {\bibfield
  {journal} {\bibinfo  {journal} {Chem. Mater.}\ }\textbf {\bibinfo {volume}
  {26}},\ \bibinfo {pages} {5042} (\bibinfo {year} {2014})}\BibitemShut
  {NoStop}%
\bibitem [{\citenamefont {Lange}\ \emph {et~al.}(2015)\citenamefont {Lange},
  \citenamefont {Reiter}, \citenamefont {Kniepert}, \citenamefont {Piersimoni},
  \citenamefont {P{\"a}tzel}, \citenamefont {Hildebrandt}, \citenamefont
  {Brenner}, \citenamefont {Hecht},\ and\ \citenamefont
  {Neher}}]{lange2015zinc}%
  \BibitemOpen
  \bibfield  {author} {\bibinfo {author} {\bibfnamefont {I.}~\bibnamefont
  {Lange}}, \bibinfo {author} {\bibfnamefont {S.}~\bibnamefont {Reiter}},
  \bibinfo {author} {\bibfnamefont {J.}~\bibnamefont {Kniepert}}, \bibinfo
  {author} {\bibfnamefont {F.}~\bibnamefont {Piersimoni}}, \bibinfo {author}
  {\bibfnamefont {M.}~\bibnamefont {P{\"a}tzel}}, \bibinfo {author}
  {\bibfnamefont {J.}~\bibnamefont {Hildebrandt}}, \bibinfo {author}
  {\bibfnamefont {T.}~\bibnamefont {Brenner}}, \bibinfo {author} {\bibfnamefont
  {S.}~\bibnamefont {Hecht}}, \ and\ \bibinfo {author} {\bibfnamefont
  {D.}~\bibnamefont {Neher}},\ }\href@noop {} {\bibfield  {journal} {\bibinfo
  {journal} {Appl. Phys. Lett.}\ }\textbf {\bibinfo {volume} {106}},\ \bibinfo
  {pages} {32\_1} (\bibinfo {year} {2015})}\BibitemShut {NoStop}%
\bibitem [{\citenamefont {Friede}\ \emph {et~al.}(2015)\citenamefont {Friede},
  \citenamefont {Kuehn}, \citenamefont {Sadofev}, \citenamefont {Blumstengel},
  \citenamefont {Henneberger},\ and\ \citenamefont
  {Elsaesser}}]{friede2015nanoscale}%
  \BibitemOpen
  \bibfield  {author} {\bibinfo {author} {\bibfnamefont {S.}~\bibnamefont
  {Friede}}, \bibinfo {author} {\bibfnamefont {S.}~\bibnamefont {Kuehn}},
  \bibinfo {author} {\bibfnamefont {S.}~\bibnamefont {Sadofev}}, \bibinfo
  {author} {\bibfnamefont {S.}~\bibnamefont {Blumstengel}}, \bibinfo {author}
  {\bibfnamefont {F.}~\bibnamefont {Henneberger}}, \ and\ \bibinfo {author}
  {\bibfnamefont {T.}~\bibnamefont {Elsaesser}},\ }\href@noop {} {\bibfield
  {journal} {\bibinfo  {journal} {Phys. Rev. B}\ }\textbf {\bibinfo {volume}
  {91}},\ \bibinfo {pages} {121415} (\bibinfo {year} {2015})}\BibitemShut
  {NoStop}%
\bibitem [{\citenamefont {Schlesinger}\ \emph {et~al.}(2015)\citenamefont
  {Schlesinger}, \citenamefont {Bianchi}, \citenamefont {Blumstengel},
  \citenamefont {Christodoulou}, \citenamefont {Ovsyannikov}, \citenamefont
  {Kobin}, \citenamefont {Moudgil}, \citenamefont {Barlow}, \citenamefont
  {Hecht}, \citenamefont {Marder} \emph {et~al.}}]{schlesinger2015efficient}%
  \BibitemOpen
  \bibfield  {author} {\bibinfo {author} {\bibfnamefont {R.}~\bibnamefont
  {Schlesinger}}, \bibinfo {author} {\bibfnamefont {F.}~\bibnamefont
  {Bianchi}}, \bibinfo {author} {\bibfnamefont {S.}~\bibnamefont
  {Blumstengel}}, \bibinfo {author} {\bibfnamefont {C.}~\bibnamefont
  {Christodoulou}}, \bibinfo {author} {\bibfnamefont {R.}~\bibnamefont
  {Ovsyannikov}}, \bibinfo {author} {\bibfnamefont {B.}~\bibnamefont {Kobin}},
  \bibinfo {author} {\bibfnamefont {K.}~\bibnamefont {Moudgil}}, \bibinfo
  {author} {\bibfnamefont {S.}~\bibnamefont {Barlow}}, \bibinfo {author}
  {\bibfnamefont {S.}~\bibnamefont {Hecht}}, \bibinfo {author} {\bibfnamefont
  {S.}~\bibnamefont {Marder}},  \emph {et~al.},\ }\href@noop {} {\bibfield
  {journal} {\bibinfo  {journal} {Nat Commun.}\ }\textbf {\bibinfo {volume}
  {6}},\ \bibinfo {pages} {6754} (\bibinfo {year} {2015})}\BibitemShut
  {NoStop}%
\bibitem [{\citenamefont {Timpel}\ \emph {et~al.}(2015)\citenamefont {Timpel},
  \citenamefont {Nardi}, \citenamefont {Ligorio}, \citenamefont {Wegner},
  \citenamefont {Pätzel}, \citenamefont {Kobin}, \citenamefont {Hecht},\ and\
  \citenamefont {Koch}}]{timpel2015energy}%
  \BibitemOpen
  \bibfield  {author} {\bibinfo {author} {\bibfnamefont {M.}~\bibnamefont
  {Timpel}}, \bibinfo {author} {\bibfnamefont {M.~V.}\ \bibnamefont {Nardi}},
  \bibinfo {author} {\bibfnamefont {G.}~\bibnamefont {Ligorio}}, \bibinfo
  {author} {\bibfnamefont {B.}~\bibnamefont {Wegner}}, \bibinfo {author}
  {\bibfnamefont {M.}~\bibnamefont {Pätzel}}, \bibinfo {author}
  {\bibfnamefont {B.}~\bibnamefont {Kobin}}, \bibinfo {author} {\bibfnamefont
  {S.}~\bibnamefont {Hecht}}, \ and\ \bibinfo {author} {\bibfnamefont
  {N.}~\bibnamefont {Koch}},\ }\href@noop {} {\bibfield  {journal} {\bibinfo
  {journal} {ACS Appl. Mater. Interfaces}\ }\textbf {\bibinfo {volume} {7}},\
  \bibinfo {pages} {11900} (\bibinfo {year} {2015})}\BibitemShut {NoStop}%
\bibitem [{\citenamefont {Kirmse}\ \emph {et~al.}(2016)\citenamefont {Kirmse},
  \citenamefont {Sparenberg}, \citenamefont {Zykov}, \citenamefont {Sadofev},
  \citenamefont {Kowarik},\ and\ \citenamefont
  {Blumstengel}}]{kirmse2016structure}%
  \BibitemOpen
  \bibfield  {author} {\bibinfo {author} {\bibfnamefont {H.}~\bibnamefont
  {Kirmse}}, \bibinfo {author} {\bibfnamefont {M.}~\bibnamefont {Sparenberg}},
  \bibinfo {author} {\bibfnamefont {A.}~\bibnamefont {Zykov}}, \bibinfo
  {author} {\bibfnamefont {S.}~\bibnamefont {Sadofev}}, \bibinfo {author}
  {\bibfnamefont {S.}~\bibnamefont {Kowarik}}, \ and\ \bibinfo {author}
  {\bibfnamefont {S.}~\bibnamefont {Blumstengel}},\ }\href@noop {} {\bibfield
  {journal} {\bibinfo  {journal} {Cryst. Growth Des.}\ }\textbf {\bibinfo
  {volume} {16}},\ \bibinfo {pages} {2789} (\bibinfo {year}
  {2016})}\BibitemShut {NoStop}%
\bibitem [{\citenamefont {Hofmann}\ and\ \citenamefont
  {Rinke}(2017)}]{hofmann2017band}%
  \BibitemOpen
  \bibfield  {author} {\bibinfo {author} {\bibfnamefont {O.~T.}\ \bibnamefont
  {Hofmann}}\ and\ \bibinfo {author} {\bibfnamefont {P.}~\bibnamefont
  {Rinke}},\ }\href@noop {} {\bibfield  {journal} {\bibinfo  {journal} {Adv.
  Electron. Mater.}\ }\textbf {\bibinfo {volume} {3}},\ \bibinfo {pages}
  {1600373} (\bibinfo {year} {2017})}\BibitemShut {NoStop}%
\bibitem [{\citenamefont {St{\"a}hler}\ and\ \citenamefont
  {Rinke}(2017)}]{stahler2017global}%
  \BibitemOpen
  \bibfield  {author} {\bibinfo {author} {\bibfnamefont {J.}~\bibnamefont
  {St{\"a}hler}}\ and\ \bibinfo {author} {\bibfnamefont {P.}~\bibnamefont
  {Rinke}},\ }\href@noop {} {\bibfield  {journal} {\bibinfo  {journal} {Chem.
  Phys.}\ }\textbf {\bibinfo {volume} {485}},\ \bibinfo {pages} {149} (\bibinfo
  {year} {2017})}\BibitemShut {NoStop}%
\bibitem [{\citenamefont {Wang}\ \emph {et~al.}(2018)\citenamefont {Wang},
  \citenamefont {Ligorio}, \citenamefont {Diez-Cabanes}, \citenamefont
  {Cornil}, \citenamefont {Kobin}, \citenamefont {Hildebrandt}, \citenamefont
  {Nardi}, \citenamefont {Timpel}, \citenamefont {Hecht}, \citenamefont
  {Cornil} \emph {et~al.}}]{wang2018dynamic}%
  \BibitemOpen
  \bibfield  {author} {\bibinfo {author} {\bibfnamefont {Q.}~\bibnamefont
  {Wang}}, \bibinfo {author} {\bibfnamefont {G.}~\bibnamefont {Ligorio}},
  \bibinfo {author} {\bibfnamefont {V.}~\bibnamefont {Diez-Cabanes}}, \bibinfo
  {author} {\bibfnamefont {D.}~\bibnamefont {Cornil}}, \bibinfo {author}
  {\bibfnamefont {B.}~\bibnamefont {Kobin}}, \bibinfo {author} {\bibfnamefont
  {J.}~\bibnamefont {Hildebrandt}}, \bibinfo {author} {\bibfnamefont {M.~V.}\
  \bibnamefont {Nardi}}, \bibinfo {author} {\bibfnamefont {M.}~\bibnamefont
  {Timpel}}, \bibinfo {author} {\bibfnamefont {S.}~\bibnamefont {Hecht}},
  \bibinfo {author} {\bibfnamefont {J.}~\bibnamefont {Cornil}},  \emph
  {et~al.},\ }\href@noop {} {\bibfield  {journal} {\bibinfo  {journal} {Adv.
  Funct. Mater.}\ ,\ \bibinfo {pages} {1800716}} (\bibinfo {year}
  {2018})}\BibitemShut {NoStop}%
\bibitem [{\citenamefont {Hill}\ \emph {et~al.}(1998)\citenamefont {Hill},
  \citenamefont {Rajagopal}, \citenamefont {Kahn},\ and\ \citenamefont
  {Hu}}]{hill1998molecular}%
  \BibitemOpen
  \bibfield  {author} {\bibinfo {author} {\bibfnamefont {I.}~\bibnamefont
  {Hill}}, \bibinfo {author} {\bibfnamefont {A.}~\bibnamefont {Rajagopal}},
  \bibinfo {author} {\bibfnamefont {A.}~\bibnamefont {Kahn}}, \ and\ \bibinfo
  {author} {\bibfnamefont {Y.}~\bibnamefont {Hu}},\ }\href@noop {} {\bibfield
  {journal} {\bibinfo  {journal} {Appl. Phys. Lett.}\ }\textbf {\bibinfo
  {volume} {73}},\ \bibinfo {pages} {662} (\bibinfo {year} {1998})}\BibitemShut
  {NoStop}%
\bibitem [{\citenamefont {Hofmann}\ \emph {et~al.}(2013)\citenamefont
  {Hofmann}, \citenamefont {Deinert}, \citenamefont {Xu}, \citenamefont
  {Rinke}, \citenamefont {St{\"a}hler}, \citenamefont {Wolf},\ and\
  \citenamefont {Scheffler}}]{Hofmann}%
  \BibitemOpen
  \bibfield  {author} {\bibinfo {author} {\bibfnamefont {O.~T.}\ \bibnamefont
  {Hofmann}}, \bibinfo {author} {\bibfnamefont {J.-C.}\ \bibnamefont
  {Deinert}}, \bibinfo {author} {\bibfnamefont {Y.}~\bibnamefont {Xu}},
  \bibinfo {author} {\bibfnamefont {P.}~\bibnamefont {Rinke}}, \bibinfo
  {author} {\bibfnamefont {J.}~\bibnamefont {St{\"a}hler}}, \bibinfo {author}
  {\bibfnamefont {M.}~\bibnamefont {Wolf}}, \ and\ \bibinfo {author}
  {\bibfnamefont {M.}~\bibnamefont {Scheffler}},\ }\href@noop {} {\bibfield
  {journal} {\bibinfo  {journal} {J. Chem. Phys.}\ }\textbf {\bibinfo {volume}
  {139}},\ \bibinfo {pages} {174701} (\bibinfo {year} {2013})}\BibitemShut
  {NoStop}%
\bibitem [{\citenamefont {Draxl}\ \emph {et~al.}(2014)\citenamefont {Draxl},
  \citenamefont {Nabok},\ and\ \citenamefont {Hannewald}}]{DNH}%
  \BibitemOpen
  \bibfield  {author} {\bibinfo {author} {\bibfnamefont {C.}~\bibnamefont
  {Draxl}}, \bibinfo {author} {\bibfnamefont {D.}~\bibnamefont {Nabok}}, \ and\
  \bibinfo {author} {\bibfnamefont {K.}~\bibnamefont {Hannewald}},\ }\href@noop
  {} {\bibfield  {journal} {\bibinfo  {journal} {Acc. Chem. Res.}\ }\textbf
  {\bibinfo {volume} {47}},\ \bibinfo {pages} {3225} (\bibinfo {year}
  {2014})}\BibitemShut {NoStop}%
\bibitem [{\citenamefont {Migani}\ \emph {et~al.}(2014)\citenamefont {Migani},
  \citenamefont {Mowbray}, \citenamefont {Zhao}, \citenamefont {Petek},\ and\
  \citenamefont {Rubio}}]{migani2014quasiparticle}%
  \BibitemOpen
  \bibfield  {author} {\bibinfo {author} {\bibfnamefont {A.}~\bibnamefont
  {Migani}}, \bibinfo {author} {\bibfnamefont {D.~J.}\ \bibnamefont {Mowbray}},
  \bibinfo {author} {\bibfnamefont {J.}~\bibnamefont {Zhao}}, \bibinfo {author}
  {\bibfnamefont {H.}~\bibnamefont {Petek}}, \ and\ \bibinfo {author}
  {\bibfnamefont {A.}~\bibnamefont {Rubio}},\ }\href@noop {} {\bibfield
  {journal} {\bibinfo  {journal} {J. Chem. Theory Comput.}\ }\textbf {\bibinfo
  {volume} {10}},\ \bibinfo {pages} {2103} (\bibinfo {year}
  {2014})}\BibitemShut {NoStop}%
\bibitem [{\citenamefont {Liu}\ \emph {et~al.}(2017)\citenamefont {Liu},
  \citenamefont {Egger}, \citenamefont {Refaely-Abramson}, \citenamefont
  {Kronik},\ and\ \citenamefont {Neaton}}]{liu2017energy}%
  \BibitemOpen
  \bibfield  {author} {\bibinfo {author} {\bibfnamefont {Z.-F.}\ \bibnamefont
  {Liu}}, \bibinfo {author} {\bibfnamefont {D.~A.}\ \bibnamefont {Egger}},
  \bibinfo {author} {\bibfnamefont {S.}~\bibnamefont {Refaely-Abramson}},
  \bibinfo {author} {\bibfnamefont {L.}~\bibnamefont {Kronik}}, \ and\ \bibinfo
  {author} {\bibfnamefont {J.~B.}\ \bibnamefont {Neaton}},\ }\href@noop {}
  {\bibfield  {journal} {\bibinfo  {journal} {J. Chem. Phys.}\ }\textbf
  {\bibinfo {volume} {146}},\ \bibinfo {pages} {092326} (\bibinfo {year}
  {2017})}\BibitemShut {NoStop}%
\bibitem [{\citenamefont {Milko}\ \emph {et~al.}(2013)\citenamefont {Milko},
  \citenamefont {~}, \citenamefont {Blondeau}, \citenamefont {Menna},
  \citenamefont {Gao}, \citenamefont {Loi},\ and\ \citenamefont
  {Draxl}}]{milko2013evidence}%
  \BibitemOpen
  \bibfield  {author} {\bibinfo {author} {\bibfnamefont {M.}~\bibnamefont
  {Milko}}, \bibinfo {author} {\bibfnamefont {P.}~\bibnamefont {~}}, \bibinfo
  {author} {\bibfnamefont {P.}~\bibnamefont {Blondeau}}, \bibinfo {author}
  {\bibfnamefont {E.}~\bibnamefont {Menna}}, \bibinfo {author} {\bibfnamefont
  {J.}~\bibnamefont {Gao}}, \bibinfo {author} {\bibfnamefont {M.~A.}\
  \bibnamefont {Loi}}, \ and\ \bibinfo {author} {\bibfnamefont
  {C.}~\bibnamefont {Draxl}},\ }\href@noop {} {\bibfield  {journal} {\bibinfo
  {journal} {J. Phys. Chem. Lett.}\ }\textbf {\bibinfo {volume} {4}},\ \bibinfo
  {pages} {2664} (\bibinfo {year} {2013})}\BibitemShut {NoStop}%
\bibitem [{\citenamefont {Mowbray}\ and\ \citenamefont
  {Migani}(2016)}]{mowbray2016optical}%
  \BibitemOpen
  \bibfield  {author} {\bibinfo {author} {\bibfnamefont {D.~J.}\ \bibnamefont
  {Mowbray}}\ and\ \bibinfo {author} {\bibfnamefont {A.}~\bibnamefont
  {Migani}},\ }\href@noop {} {\bibfield  {journal} {\bibinfo  {journal} {J.
  Chem. Theory Comput.}\ }\textbf {\bibinfo {volume} {12}},\ \bibinfo {pages}
  {2843} (\bibinfo {year} {2016})}\BibitemShut {NoStop}%
\bibitem [{\citenamefont {Fu}\ \emph {et~al.}(2017)\citenamefont {Fu},
  \citenamefont {Cocchi}, \citenamefont {Nabok}, \citenamefont {Gulans},\ and\
  \citenamefont {Draxl}}]{fu2017graphene}%
  \BibitemOpen
  \bibfield  {author} {\bibinfo {author} {\bibfnamefont {Q.}~\bibnamefont
  {Fu}}, \bibinfo {author} {\bibfnamefont {C.}~\bibnamefont {Cocchi}}, \bibinfo
  {author} {\bibfnamefont {D.}~\bibnamefont {Nabok}}, \bibinfo {author}
  {\bibfnamefont {A.}~\bibnamefont {Gulans}}, \ and\ \bibinfo {author}
  {\bibfnamefont {C.}~\bibnamefont {Draxl}},\ }\href@noop {} {\bibfield
  {journal} {\bibinfo  {journal} {Phys. Chem. Chem. Phys.}\ }\textbf {\bibinfo
  {volume} {19}},\ \bibinfo {pages} {6196} (\bibinfo {year}
  {2017})}\BibitemShut {NoStop}%
\bibitem [{\citenamefont {Ljungberg}\ \emph {et~al.}(2017)\citenamefont
  {Ljungberg}, \citenamefont {V{\"a}nsk{\"a}}, \citenamefont {Koval},
  \citenamefont {Koch}, \citenamefont {Kira},\ and\ \citenamefont
  {S{\'a}nchez-Portal}}]{ljungberg2017charge}%
  \BibitemOpen
  \bibfield  {author} {\bibinfo {author} {\bibfnamefont {M.}~\bibnamefont
  {Ljungberg}}, \bibinfo {author} {\bibfnamefont {O.}~\bibnamefont
  {V{\"a}nsk{\"a}}}, \bibinfo {author} {\bibfnamefont {P.}~\bibnamefont
  {Koval}}, \bibinfo {author} {\bibfnamefont {S.}~\bibnamefont {Koch}},
  \bibinfo {author} {\bibfnamefont {M.}~\bibnamefont {Kira}}, \ and\ \bibinfo
  {author} {\bibfnamefont {D.}~\bibnamefont {S{\'a}nchez-Portal}},\ }\href@noop
  {} {\bibfield  {journal} {\bibinfo  {journal} {New J. Phys.}\ }\textbf
  {\bibinfo {volume} {19}},\ \bibinfo {pages} {033019} (\bibinfo {year}
  {2017})}\BibitemShut {NoStop}%
\bibitem [{\citenamefont {Perdew}\ and\ \citenamefont {Wang}(1992)}]{LDA_PW}%
  \BibitemOpen
  \bibfield  {author} {\bibinfo {author} {\bibfnamefont {J.~P.}\ \bibnamefont
  {Perdew}}\ and\ \bibinfo {author} {\bibfnamefont {Y.}~\bibnamefont {Wang}},\
  }\href {\doibase 10.1103/PhysRevB.45.13244} {\bibfield  {journal} {\bibinfo
  {journal} {Phys. Rev. B}\ }\textbf {\bibinfo {volume} {45}},\ \bibinfo
  {pages} {13244} (\bibinfo {year} {1992})}\BibitemShut {NoStop}%
\bibitem [{\citenamefont {Ceperley}\ and\ \citenamefont
  {Alder}(1980)}]{LDA_CA}%
  \BibitemOpen
  \bibfield  {author} {\bibinfo {author} {\bibfnamefont {D.~M.}\ \bibnamefont
  {Ceperley}}\ and\ \bibinfo {author} {\bibfnamefont {B.~J.}\ \bibnamefont
  {Alder}},\ }\href {\doibase 10.1103/PhysRevLett.45.566} {\bibfield  {journal}
  {\bibinfo  {journal} {Phys. Rev. Lett.}\ }\textbf {\bibinfo {volume} {45}},\
  \bibinfo {pages} {566} (\bibinfo {year} {1980})}\BibitemShut {NoStop}%
\bibitem [{\citenamefont {Gulans}\ \emph {et~al.}(2014)\citenamefont {Gulans},
  \citenamefont {Kontur}, \citenamefont {Meisenbichler}, \citenamefont {Nabok},
  \citenamefont {Pavone}, \citenamefont {Rigamonti}, \citenamefont
  {Sagmeister}, \citenamefont {Werner},\ and\ \citenamefont
  {Draxl}}]{exciting}%
  \BibitemOpen
  \bibfield  {author} {\bibinfo {author} {\bibfnamefont {A.}~\bibnamefont
  {Gulans}}, \bibinfo {author} {\bibfnamefont {S.}~\bibnamefont {Kontur}},
  \bibinfo {author} {\bibfnamefont {C.}~\bibnamefont {Meisenbichler}}, \bibinfo
  {author} {\bibfnamefont {D.}~\bibnamefont {Nabok}}, \bibinfo {author}
  {\bibfnamefont {P.}~\bibnamefont {Pavone}}, \bibinfo {author} {\bibfnamefont
  {S.}~\bibnamefont {Rigamonti}}, \bibinfo {author} {\bibfnamefont
  {S.}~\bibnamefont {Sagmeister}}, \bibinfo {author} {\bibfnamefont
  {U.}~\bibnamefont {Werner}}, \ and\ \bibinfo {author} {\bibfnamefont
  {C.}~\bibnamefont {Draxl}},\ }\href@noop {} {\bibfield  {journal} {\bibinfo
  {journal} {J. Phys. Condens. Matter}\ }\textbf {\bibinfo {volume} {26}},\
  \bibinfo {pages} {363202} (\bibinfo {year} {2014})}\BibitemShut {NoStop}%
\bibitem [{\citenamefont {Momma}\ and\ \citenamefont {Izumi}(2011)}]{Vesta}%
  \BibitemOpen
  \bibfield  {author} {\bibinfo {author} {\bibfnamefont {K.}~\bibnamefont
  {Momma}}\ and\ \bibinfo {author} {\bibfnamefont {F.}~\bibnamefont {Izumi}},\
  }\href@noop {} {\bibfield  {journal} {\bibinfo  {journal} {J. Appl.
  Crystallogr.}\ }\textbf {\bibinfo {volume} {44}},\ \bibinfo {pages} {1272}
  (\bibinfo {year} {2011})}\BibitemShut {NoStop}%
\bibitem [{\citenamefont {Neaton}\ \emph {et~al.}(2006)\citenamefont {Neaton},
  \citenamefont {Hybertsen},\ and\ \citenamefont {Louie}}]{Polarization1}%
  \BibitemOpen
  \bibfield  {author} {\bibinfo {author} {\bibfnamefont {J.~B.}\ \bibnamefont
  {Neaton}}, \bibinfo {author} {\bibfnamefont {M.~S.}\ \bibnamefont
  {Hybertsen}}, \ and\ \bibinfo {author} {\bibfnamefont {S.~G.}\ \bibnamefont
  {Louie}},\ }\href {\doibase 10.1103/PhysRevLett.97.216405} {\bibfield
  {journal} {\bibinfo  {journal} {Phys. Rev. Lett.}\ }\textbf {\bibinfo
  {volume} {97}},\ \bibinfo {pages} {216405} (\bibinfo {year}
  {2006})}\BibitemShut {NoStop}%
\bibitem [{\citenamefont {Refaely-Abramson}\ \emph {et~al.}(2013)\citenamefont
  {Refaely-Abramson}, \citenamefont {Sharifzadeh}, \citenamefont {Jain},
  \citenamefont {Baer}, \citenamefont {Neaton},\ and\ \citenamefont
  {Kronik}}]{Polarization2}%
  \BibitemOpen
  \bibfield  {author} {\bibinfo {author} {\bibfnamefont {S.}~\bibnamefont
  {Refaely-Abramson}}, \bibinfo {author} {\bibfnamefont {S.}~\bibnamefont
  {Sharifzadeh}}, \bibinfo {author} {\bibfnamefont {M.}~\bibnamefont {Jain}},
  \bibinfo {author} {\bibfnamefont {R.}~\bibnamefont {Baer}}, \bibinfo {author}
  {\bibfnamefont {J.~B.}\ \bibnamefont {Neaton}}, \ and\ \bibinfo {author}
  {\bibfnamefont {L.}~\bibnamefont {Kronik}},\ }\href {\doibase
  10.1103/PhysRevB.88.081204} {\bibfield  {journal} {\bibinfo  {journal} {Phys.
  Rev. B}\ }\textbf {\bibinfo {volume} {88}},\ \bibinfo {pages} {081204}
  (\bibinfo {year} {2013})}\BibitemShut {NoStop}%
\bibitem [{\citenamefont {Despoja}\ \emph {et~al.}(2013)\citenamefont
  {Despoja}, \citenamefont {Lon\ifmmode \check{c}\else
  \v{c}\fi{}ari\ifmmode~\acute{c}\else \'{c}\fi{}}, \citenamefont {Mowbray},\
  and\ \citenamefont {Maru\ifmmode \check{s}\else
  \v{s}\fi{}i\ifmmode~\acute{c}\else \'{c}\fi{}}}]{Polarization3}%
  \BibitemOpen
  \bibfield  {author} {\bibinfo {author} {\bibfnamefont {V.}~\bibnamefont
  {Despoja}}, \bibinfo {author} {\bibfnamefont {I.}~\bibnamefont {Lon\ifmmode
  \check{c}\else \v{c}\fi{}ari\ifmmode~\acute{c}\else \'{c}\fi{}}}, \bibinfo
  {author} {\bibfnamefont {D.~J.}\ \bibnamefont {Mowbray}}, \ and\ \bibinfo
  {author} {\bibfnamefont {L.}~\bibnamefont {Maru\ifmmode \check{s}\else
  \v{s}\fi{}i\ifmmode~\acute{c}\else \'{c}\fi{}}},\ }\href {\doibase
  10.1103/PhysRevB.88.235437} {\bibfield  {journal} {\bibinfo  {journal} {Phys.
  Rev. B}\ }\textbf {\bibinfo {volume} {88}},\ \bibinfo {pages} {235437}
  (\bibinfo {year} {2013})}\BibitemShut {NoStop}%
\bibitem [{\citenamefont {Garcia-Lastra}\ \emph {et~al.}(2009)\citenamefont
  {Garcia-Lastra}, \citenamefont {Rostgaard}, \citenamefont {Rubio},\ and\
  \citenamefont {Thygesen}}]{Polarization4}%
  \BibitemOpen
  \bibfield  {author} {\bibinfo {author} {\bibfnamefont {J.~M.}\ \bibnamefont
  {Garcia-Lastra}}, \bibinfo {author} {\bibfnamefont {C.}~\bibnamefont
  {Rostgaard}}, \bibinfo {author} {\bibfnamefont {A.}~\bibnamefont {Rubio}}, \
  and\ \bibinfo {author} {\bibfnamefont {K.~S.}\ \bibnamefont {Thygesen}},\
  }\href {\doibase 10.1103/PhysRevB.80.245427} {\bibfield  {journal} {\bibinfo
  {journal} {Phys. Rev. B}\ }\textbf {\bibinfo {volume} {80}},\ \bibinfo
  {pages} {245427} (\bibinfo {year} {2009})}\BibitemShut {NoStop}%
\bibitem [{\citenamefont {Egger}\ \emph {et~al.}(2015)\citenamefont {Egger},
  \citenamefont {Liu}, \citenamefont {Neaton},\ and\ \citenamefont
  {Kronik}}]{egger2015reliable}%
  \BibitemOpen
  \bibfield  {author} {\bibinfo {author} {\bibfnamefont {D.~A.}\ \bibnamefont
  {Egger}}, \bibinfo {author} {\bibfnamefont {Z.-F.}\ \bibnamefont {Liu}},
  \bibinfo {author} {\bibfnamefont {J.~B.}\ \bibnamefont {Neaton}}, \ and\
  \bibinfo {author} {\bibfnamefont {L.}~\bibnamefont {Kronik}},\ }\href@noop {}
  {\bibfield  {journal} {\bibinfo  {journal} {Nano lett.}\ }\textbf {\bibinfo
  {volume} {15}},\ \bibinfo {pages} {2448} (\bibinfo {year}
  {2015})}\BibitemShut {NoStop}%
\bibitem [{\citenamefont {Gori}\ \emph {et~al.}(2010)\citenamefont {Gori},
  \citenamefont {Rakel}, \citenamefont {Cobet}, \citenamefont {Richter},
  \citenamefont {Esser}, \citenamefont {Hoffmann}, \citenamefont {Del~Sole},
  \citenamefont {Cricenti},\ and\ \citenamefont {Pulci}}]{gori2010optical}%
  \BibitemOpen
  \bibfield  {author} {\bibinfo {author} {\bibfnamefont {P.}~\bibnamefont
  {Gori}}, \bibinfo {author} {\bibfnamefont {M.}~\bibnamefont {Rakel}},
  \bibinfo {author} {\bibfnamefont {C.}~\bibnamefont {Cobet}}, \bibinfo
  {author} {\bibfnamefont {W.}~\bibnamefont {Richter}}, \bibinfo {author}
  {\bibfnamefont {N.}~\bibnamefont {Esser}}, \bibinfo {author} {\bibfnamefont
  {A.}~\bibnamefont {Hoffmann}}, \bibinfo {author} {\bibfnamefont
  {R.}~\bibnamefont {Del~Sole}}, \bibinfo {author} {\bibfnamefont
  {A.}~\bibnamefont {Cricenti}}, \ and\ \bibinfo {author} {\bibfnamefont
  {O.}~\bibnamefont {Pulci}},\ }\href@noop {} {\bibfield  {journal} {\bibinfo
  {journal} {Phys. Rev. B}\ }\textbf {\bibinfo {volume} {81}},\ \bibinfo
  {pages} {125207} (\bibinfo {year} {2010})}\BibitemShut {NoStop}%
\bibitem [{\citenamefont {Puschnig}\ and\ \citenamefont
  {Ambrosch-Draxl}(2002)}]{Puschnig2002}%
  \BibitemOpen
  \bibfield  {author} {\bibinfo {author} {\bibfnamefont {P.}~\bibnamefont
  {Puschnig}}\ and\ \bibinfo {author} {\bibfnamefont {C.}~\bibnamefont
  {Ambrosch-Draxl}},\ }\href {\doibase 10.1103/PhysRevLett.89.056405}
  {\bibfield  {journal} {\bibinfo  {journal} {Phys. Rev. Lett.}\ }\textbf
  {\bibinfo {volume} {89}},\ \bibinfo {pages} {056405} (\bibinfo {year}
  {2002})}\BibitemShut {NoStop}%
\bibitem [{\citenamefont {Deinert}\ \emph {et~al.}(2014)\citenamefont
  {Deinert}, \citenamefont {Wegkamp}, \citenamefont {Meyer}, \citenamefont
  {Richter}, \citenamefont {Wolf},\ and\ \citenamefont
  {St{\"a}hler}}]{deinert2014ultrafast}%
  \BibitemOpen
  \bibfield  {author} {\bibinfo {author} {\bibfnamefont {J.-C.}\ \bibnamefont
  {Deinert}}, \bibinfo {author} {\bibfnamefont {D.}~\bibnamefont {Wegkamp}},
  \bibinfo {author} {\bibfnamefont {M.}~\bibnamefont {Meyer}}, \bibinfo
  {author} {\bibfnamefont {C.}~\bibnamefont {Richter}}, \bibinfo {author}
  {\bibfnamefont {M.}~\bibnamefont {Wolf}}, \ and\ \bibinfo {author}
  {\bibfnamefont {J.}~\bibnamefont {St{\"a}hler}},\ }\href@noop {} {\bibfield
  {journal} {\bibinfo  {journal} {Phys. Rev. Lett.}\ }\textbf {\bibinfo
  {volume} {113}},\ \bibinfo {pages} {057602} (\bibinfo {year}
  {2014})}\BibitemShut {NoStop}%
\end{thebibliography}
\providecommand{\noopsort}[1]{}\providecommand{\singleletter}[1]{#1}%

\end{document}